\documentclass[12pt]{article}
\usepackage{graphicx}
\usepackage{equations}
\usepackage{axodraw}

\makeatletter
\def\@cite#1#2{{[{#1}]\if@tempswa\typeout
{IJCGA warning: optional citation argument
ignored: `#2'} \fi}}
\newcount\@tempcntc
\def\@citex[#1]#2{\if@filesw\immediate\write\@auxout{\string\citation{#2}}\fi
  \@tempcnta\z@\@tempcntb\m@ne\def\@citea{}\@cite{\@for\@citeb:=#2\do
    {\@ifundefined
       {b@\@citeb}{\@citeo\@tempcntb\m@ne\@citea\def\@citea{,}{\bf ?}\@warning
       {Citation `\@citeb' on page \thepage \space undefined}}%
    {\setbox\z@\hbox{\global\@tempcntc0\csname b@\@citeb\endcsname\relax}%
     \ifnum\@tempcntc=\z@ \@citeo\@tempcntb\m@ne
       \@citea\def\@citea{,}\hbox{\csname b@\@citeb\endcsname}%
     \else
      \advance\@tempcntb\@ne
      \ifnum\@tempcntb=\@tempcntc
      \else\advance\@tempcntb\m@ne\@citeo
      \@tempcnta\@tempcntc\@tempcntb\@tempcntc\fi\fi}}\@citeo}{#1}}
\def\@citeo{\ifnum\@tempcnta>\@tempcntb\else\@citea\def\@citea{,}%
  \ifnum\@tempcnta=\@tempcntb\the\@tempcnta\else
   {\advance\@tempcnta\@ne\ifnum\@tempcnta=\@tempcntb \else \def\@citea{--}\fi
    \advance\@tempcnta\m@ne\the\@tempcnta\@citea\the\@tempcntb}\fi\fi}
\makeatother
\newlength{\captsize}           \let\captsize=\footnotesize
\newlength{\captwidth}          
\newlength{\beforetableskip}    
\newcommand{\capt}[1]{\begin{minipage}{\captwidth}
              \let\normalsize=\captsize
              \caption[0]{#1}
              \end{minipage}\\ \vspace{\beforetableskip}}
\makeatletter
      \long\def\@makecaption#1#2{\vskip 10 \p@
      \setbox\@tempboxa\hbox{\textbf{#1:} #2}%     %% boldface
      \ifdim \wd\@tempboxa >\hsize
            \textbf{#1:} #2\par                 %% boldface
      \else
         \hbox to \hsize{\box\@tempboxa\hfil}%         %%no centering
      \fi}
\makeatother
\newcommand\longdash{\hbox{\rm{\phantom{a}---\phantom{a}}}}

\newcommand\beq{\begin{equation}}
\newcommand\eeq{\end{equation}}
\newcommand{\ben}{\begin{enumerate}}
\newcommand{\een}{\end{enumerate}}

\newenvironment{Eqnarray}%
     {\arraycolsep 0.14em\begin{eqnarray}}{\end{eqnarray}}
\newcommand{\mathbold}[1]{\mbox{\boldmath $\bf#1$}}
\def\beqa{\begin{Eqnarray}}
\def\eeqa{\end{Eqnarray}}
\def\ifmath#1{\relax\ifmmode #1\else $#1$\fi}
\def\lsim{\mathrel{\raise.3ex\hbox{$<$\kern-.75em\lower1ex\hbox{$\sim$}}}}
\def\gsim{\mathrel{\raise.3ex\hbox{$>$\kern-.75em\lower1ex\hbox{$\sim$}}}}
\def\eq#1{eq.~(\ref{#1})}
\def\fig#1{fig.~\ref{#1}}
\def\Fig#1{Fig.~\ref{#1}}

\def\Ref#1{ref.~\cite{#1}}

\def\Sec#1{Section~\ref{#1}}

\def\phm{\phantom{-}}
\newcommand{\bb}{b\bar{b}}
\newcommand{\nn}{\nu\bar{\nu}}

\newcommand{\lpm}{\ell^+\ell^-}
\newcommand{\lsdilep}{\ell^\pm\ell^\pm jj}
\newcommand{\trilep}{\ell^\pm\ell^{\prime\pm}\ell^\mp}
\def\SM{Standard Model}

\def\tanb{\tan\beta}

\def\cosbma{\cos(\beta-\alpha)}
\def\hsm{h_{\rm SM}}
\def\mhsm{m_{h_{\rm SM}}}
\def\hl{h}
\def\ha{A}
\def\hh{H}
\def\hpm{{H^\pm}}

\def\mha{m_{\ha}}
\def\mhl{m_{\hl}}
\def\mhh{m_{\hh}}
\def\mhpm{m_{\hpm}}
\def\mhmax{m_h^{\rm max}}
\def\mz{m_Z}
\def\mw{m_W}

\def\mt{m_t}

\def\ls#1{\ifmath{_{\lower1.5pt\hbox{$\scriptstyle #1$}}}}
\def\lss#1{\ifmath{^{\,\lower2.5pt\hbox{$\scriptstyle #1$}}}}

\def\mstopa{M_{\widetilde t_1}}
\def\mstopb{M_{\widetilde t_2}}
\def\nicefrac#1#2{\hbox{${#1\over #2}$}}
\def\half{\ifmath{{\textstyle{1 \over 2}}}}

\def\npb#1#2#3{{\sl Nucl. Phys. }{\bf B#1} (#2)~#3}

\def\plb#1#2#3{{\sl Phys. Lett. }{\bf B#1} (#2) #3}
\def\prd#1#2#3{{\sl Phys. Rev. }{\bf D#1} (#2)~#3}

\def\prl#1#2#3{{\sl Phys. Rev. Lett. }{\bf #1} (#2) #3}
\def\prc#1#2#3{{\sl Phys. Reports }{\bf #1} (#2)~#3}
\def\cpc#1#2#3{{\sl Comp. Phys. Commun. }{\bf #1} (#2) #3}

\def\pr#1#2#3{{\sl Phys. Reports }{\bf #1} (#2) #3}
\def\sovnp#1#2#3{{\sl Sov. J. Nucl. Phys. }{\bf #1} (#2) #3}

\def\zpc#1#2#3{{\sl Z. Phys. }{\bf C#1} (#2) #3}
\def\ptp#1#2#3{{\sl Prog.~Theor.~Phys.~}{\bf #1} (#2) #3}

\def\aop#1#2#3{{\sl Annals~Phys.~}{\bf #1} (#2) #3}
\def\epjc#1#2#3{{\sl Eur.~Phys.~J.~}{\bf C#1} (#2) #3}

\begin{document}
%%%%%%%%%%%%%%%%%%%%%%%%%%%%%%%%%%%%%%%%%%%%%%%%%%%%%%%%%%%%%%
%
\vbox{  \large
\begin{flushright}
SCIPP 02/38 \\
December, 2002 \\
hep--ph/0212136\\
\end{flushright}
\vskip1.5cm
\begin{center}
{\LARGE\bf
Higgs Theory and Phenomenology in the\\[6pt]
Standard Model and MSSM}\\[1cm]

{\Large Howard E. Haber}\\[5pt]{\it Santa Cruz Institute for Particle
Physics,  \\ University of California, Santa Cruz, CA 95064, U.S.A.}\\
%\end{center}

\vskip2cm
\thispagestyle{empty}

{\bf Abstract}\\[1pc]

\begin{minipage}{15cm}
A short review of the theory and phenomenology of Higgs bosons is
given, with focus on the Standard Model (SM) and the minimal
supersymmetric extension of the Standard Model (MSSM).  The potential
for Higgs boson discovery at the Tevatron and LHC, and precision Higgs
studies at the LHC and a future $e^+e^-$ linear collider are briefly
surveyed.  The phenomenological challenge of the approach to the
decoupling limit, where the properties of the lightest CP-even Higgs
boson of the MSSM are nearly indistinguishable from those of the SM
Higgs boson is emphasized.
\end{minipage}  \\
\vskip2.5cm
%\begin{center}
Invited talk at \\
The 10th International Conference on Supersymmetry and \\ 
Unification of Fundamental Interactions (SUSY02) \\
17--23 June 2002, DESY Hamburg, Germany\\
\end{center}
}
\vfill
\clearpage

%%%%%%%%%%%%%%%%%%%%%%%%%%%%%%%%%%%%%%%%%%%%%%%%%%%%%%%%%%%%%%%%%%%%%
%\pagestyle{plain}
\setcounter{page}{1}

\begin{center}
{\Large\bf  Higgs Theory and Phenomenology in the}\\[0.3cm]
{\Large \bf Standard Model and MSSM}\\[1cm]
{\large Howard E. Haber}\\[5pt]
{\it Santa Cruz Institute for Particle Physics  \\
   University of California, Santa Cruz, CA 95064, U.S.A.} \\
\end{center}
\vspace{.2cm}
\begin{abstract}
A short review of the theory and phenomenology of Higgs bosons is
given, with focus on the Standard Model (SM) and the minimal
supersymmetric extension of the Standard Model (MSSM).  The potential
for Higgs boson discovery at the Tevatron and LHC, and precision Higgs
studies at the LHC and a future $e^+e^-$ linear collider are briefly
surveyed.  The phenomenological challenge of the approach to the
decoupling limit, where the properties of the lightest CP-even Higgs
boson of the MSSM are nearly indistinguishable from those of the SM
Higgs boson is emphasized.
\end{abstract}

\vspace*{.3cm}

\setcounter{footnote}{0}
\renewcommand{\thefootnote}{\arabic{footnote}}
\section{Introduction}  \label{sec:intro}

\begin{figure}[t!]
\begin{center}
\unitlength1cm
\begin{picture}(15,8.3)
\put(-1.0,-1.0){\includegraphics[width=8cm]{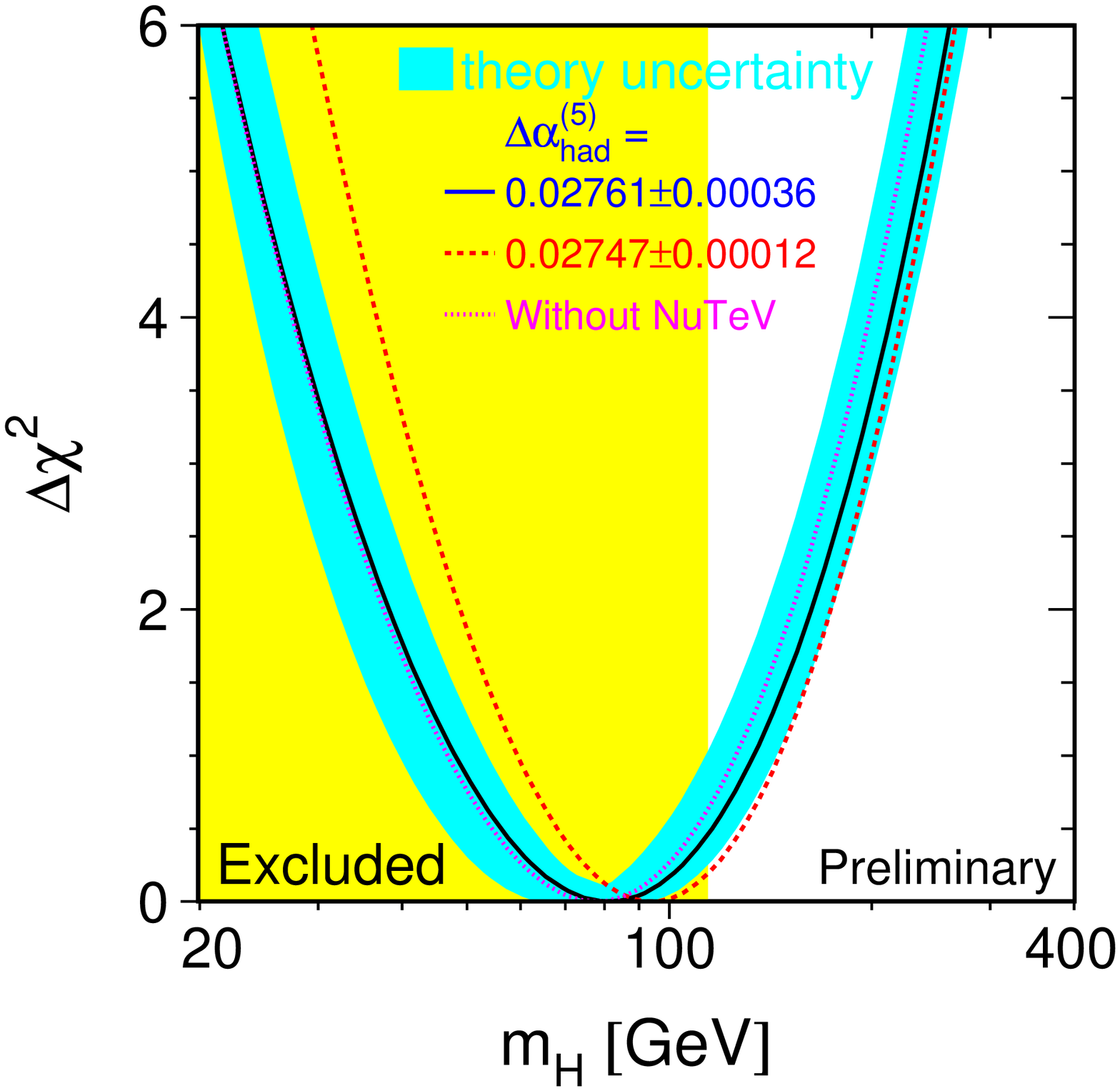}}
\put(8.05,-0.15){\includegraphics[width=7.2cm]{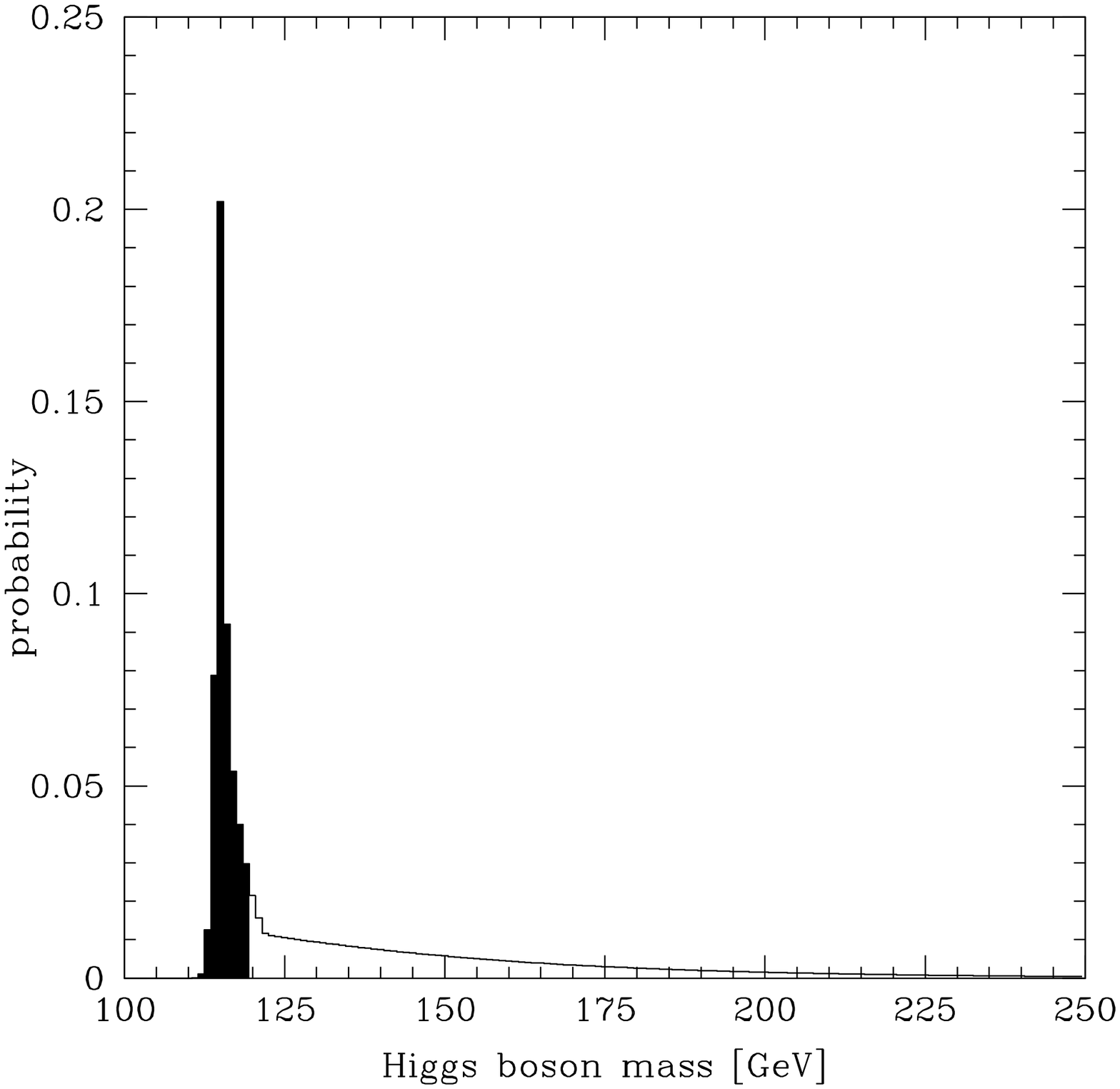}}
\end{picture}
\end{center}
\capt{\label{fig:blueband}
(a)~The ``blueband plot'' shows $\Delta
\chi^2\equiv \chi^2-\chi^2_{\min}$ as a function of the SM
Higgs mass~\protect\cite{lepewwg}.
The solid line is a result of a global fit
using all data; the band represents the theoretical
error due to missing higher order corrections.  The rectangular shaded region
shows the 95\% CL exclusion limit on the Higgs mass from direct
searches at LEP~\protect\cite{LEPHiggs}.
(b)~Probability distribution function for the Higgs boson
mass, including all available direct and indirect data~\protect\cite{erler}.
The probability is shown for 1~GeV bins.  The shaded and unshaded
regions each correspond to an integrated probability of 50\%}
\end{figure}

Despite the great successes of the LEP, SLC and Tevatron colliders
during the 1990s in verifying many detailed aspects of the Standard
Model (SM), the origin of electroweak symmetry breaking has not yet been
fully revealed.  Nevertheless, the precision electroweak data impose
some strong constraints, and seem to provide strong support for the
Standard Model with a weakly-coupled Higgs boson ($\hsm$).  The
results of the LEP Electroweak Working Group analysis shown in
\fig{fig:blueband}(a) yield \cite{lepewwg}: $\mhsm=81^{+52}_{-33}~{\rm
GeV}$, and yield a 95\%~CL upper limit of $\mhsm<193$~GeV.  These
results reflect the logarithmic sensitivity to $\mhsm$ via the
virtual Higgs loop contributions to the various electroweak
observables.  The 95\%~CL upper limit on $\mhsm$
is consistent with the direct
searches at LEP~\cite{LEPHiggs} that show no conclusive evidence for
the Higgs boson, and imply that $\mhsm> 114.4$~GeV at 95\%~CL.
\Fig{fig:blueband}(b) exhibits the most probable range of values for
the SM Higgs mass~\cite{erler}.  This mass range is consistent with a
weakly-coupled Higgs scalar that is expected to emerge from the
scalar dynamics of a self-interacting complex Higgs doublet.
% (although the Standard Model does not
%predict the mass of the Higgs boson; rather it relates it to the
%strength of the scalar self-coupling).

Based on the successes of the Standard Model global fits to
electroweak data, one can also constrain the contributions 
of new physics, which can enter through $W^\pm$ and $Z$ boson vacuum
polarization corrections.  This fact has
already served to rule out numerous models of strongly-coupled
electroweak symmetry breaking dynamics.   Nevertheless,
there are some loopholes that can be exploited to circumvent
the conclusion that the Standard Model with a light Higgs boson is
preferred.  It is possible to construct models of new physics
where the goodness of the global Standard
Model fit to precision electroweak data is not compromised
while the strong upper limit on the Higgs mass is
relaxed.  In particular, one can construct effective
operators~\cite{newoperators,murayama}
or specific models of new physics~\cite{Peskin-Wells} where
the Higgs mass is significantly larger, but the new
physics contributions to the
$W^\pm$ and $Z$ vacuum polarizations, parameterized by the
Peskin-Takeuchi~\cite{peskin} parameters $S$ and $T$,
are still consistent with the experimental data.
In addition, some have argued that the
global Standard Model fit exhibits
some internal inconsistencies~\cite{chanowitz}, which would suggest that
systematic uncertainties have been underestimated and/or new physics
beyond the Standard Model is required.
Thus, although weakly-coupled electroweak
symmetry breaking seems to be favored
by the precision electroweak data, one cannot definitively
rule out all other approaches.
However, in this review I shall assume that the 
Higgs boson is indeed weakly-coupled due to electroweak
symmetry breaking based on scalar dynamics.

\begin{figure}[t!]
\begin{center}
\includegraphics*[width=8cm]{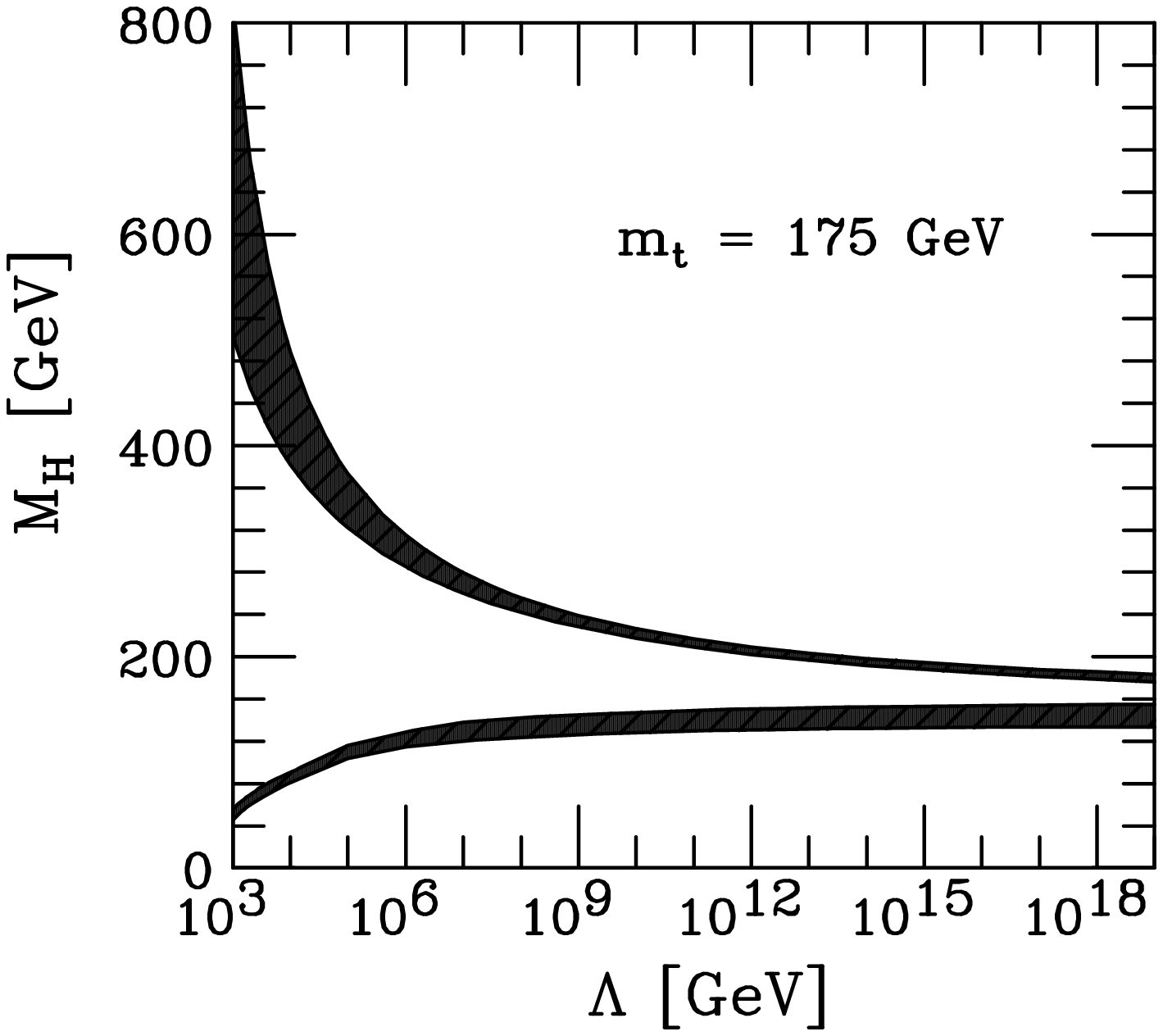}
\hspace*{5mm}
\includegraphics*[width=7cm]{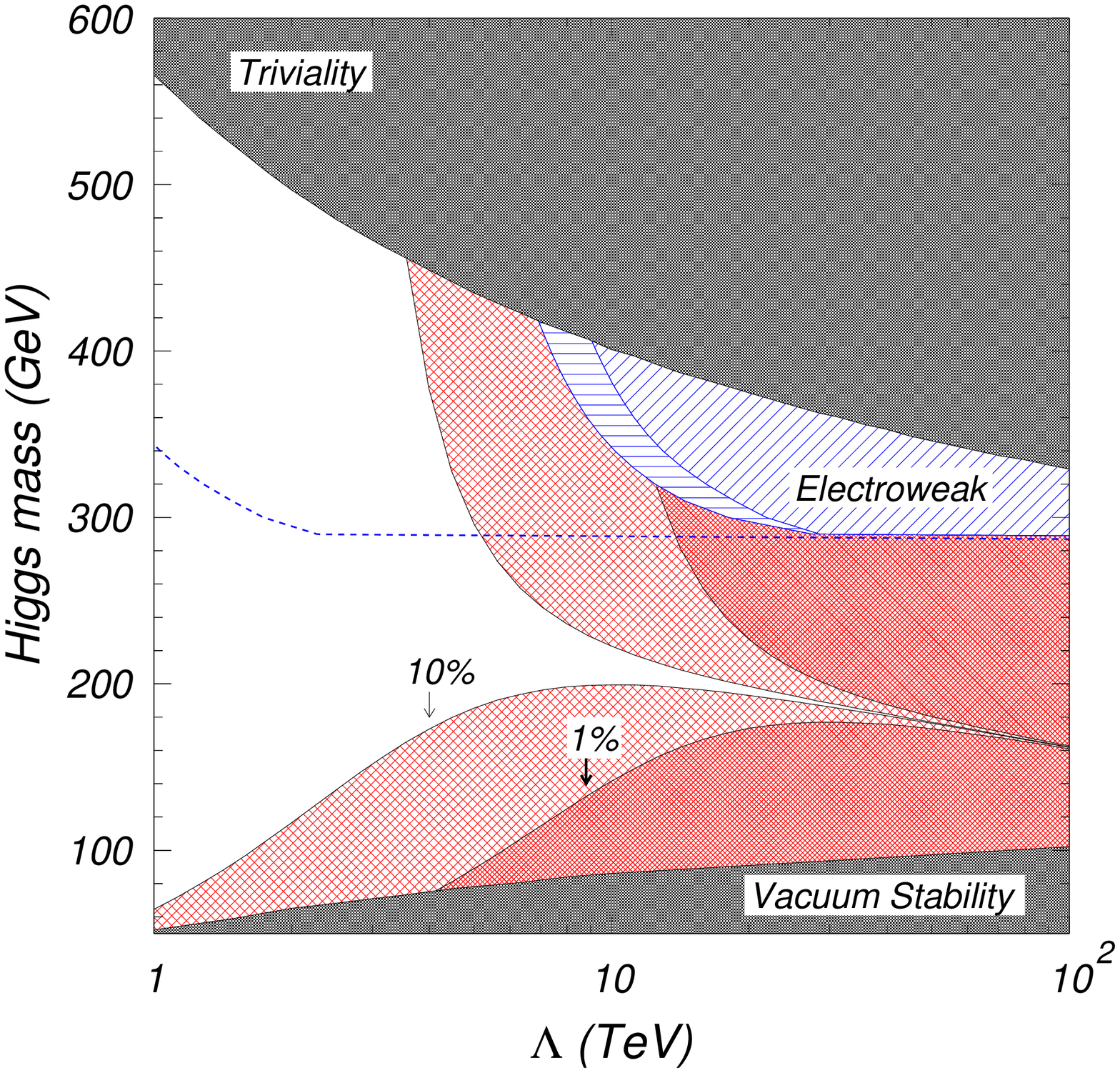}
\end{center}
\capt{\label{trivial} (a)~The upper \protect\cite{hambye}
and the lower \protect\cite{quiros}
Higgs mass bounds as a function of the
energy scale $\Lambda$ at which the Standard Model breaks down,
assuming $M_t=175$~GeV and $\alpha_s(m_Z)=0.118$, taken from
\protect\Ref{Riesselmann}.  The shaded
areas above reflect the theoretical uncertainties in the
calculations of the Higgs mass bounds.  (b)~Following \protect\Ref{murayama},
a reconsideration of the $\Lambda$ {\it vs.} Higgs mass plot with a
focus on $\Lambda<100$~TeV.  Precision electroweak measurements
restrict the parameter space to lie below the dashed line, based on a
95\% CL
fit that allows for nonzero values of $S$ and $T$ and the existence of
higher dimensional operators suppressed by $v^2/\Lambda^2$.  The
unshaded area has less than one part in ten fine-tuning.}
\end{figure}

The Standard Model is an effective field theory and provides a very good
description of the physics of fundamental particles and their
interactions at an energy scale of ${\cal O}(100)$~GeV and below.
However, there must exist some energy scale, $\Lambda$, at which the
Standard Model breaks down.
That is, the Standard Model is no longer
adequate for describing the theory above $\Lambda$, and degrees of
freedom associated with new physics
become relevant.  Although the value of $\Lambda$ is presently
unknown, the Higgs mass can provide an important constraint.
If $\mhsm$ is too large, then the
Higgs self-coupling blows up at some scale $\Lambda$ below the
Planck scale \cite{hambye}.  If $\mhsm$ is too small, then the Higgs
potential develops
a second (global) minimum at a large value of the scalar field of order
$\Lambda$ \cite{quiros}.  Thus new physics must enter at a scale
$\Lambda$
or below in order that the global minimum of the theory correspond to the
observed SU(2)$\times$U(1) broken vacuum with
$v=246$~GeV.  Given a value of
$\Lambda$, one can compute the minimum and maximum Higgs mass allowed.
The results of this computation (with shaded bands indicating the
theoretical uncertainty of the result) are illustrated in
\fig{trivial}(a)~\cite{Riesselmann}.
Consequently, a
Higgs mass range 130~GeV~$\lsim\mhsm\lsim 180$~GeV
is consistent with an effective Standard Model
that survives all the way to the Planck scale.

However, the survival of the Standard Model as an effective theory all
the way up to the Planck scale is unlikely.
Electroweak symmetry breaking dynamics driven by a weakly-coupled
elementary scalar sector requires a mechanism for the stability of
the electroweak symmetry breaking scale with respect to the
Planck scale~\cite{natural}.  However,
scalar squared-masses at one-loop order are {\it quadratically}
sensitive to $\Lambda$.   In order for this
value to be consistent with the requirement that $\mhsm\lsim\mathcal{O}(v)$,
as required from
unitarity constraints~\cite{unitarity,thacker}, 
$\Lambda\lsim {4\pi \mhsm/ g}\sim {\cal O}(1~{\rm TeV})$ (where $g$ is
an electroweak gauge coupling).
If $\Lambda$ is significantly larger than 1~TeV, then the only way
to generate a Higgs mass of $\mathcal{O}(v)$
is to have an ``unnatural'' cancellation between the bare Higgs mass
and its loop corrections.
The requirement of $\Lambda\sim\mathcal{O}(1~{\rm TeV}$) as a condition for
the absence of fine-tuning of the Higgs mass parameter
is nicely illustrated in \fig{trivial}(b)~\cite{murayama}.
%taken from \Ref{murayama}.

A viable theoretical framework that incorporates weakly-coupled Higgs
bosons and satisfies the naturalness constraint is that of weak-scale
supersymmetry \cite{susynatural}.  Since fermion masses are only
logarithmically sensitive to $\Lambda$, boson masses will exhibit the
same logarithmic sensitivity if supersymmetry is exact.  However,
supersymmetry is not an exact symmetry of fundamental particle
interactions, and thus
$\Lambda$ should be identified with the energy scale 
of supersymmetry-breaking.  Moreover, 
supersymmetry-breaking effects can induce a
radiative breaking of the electroweak symmetry due to the effects of
the large Higgs-top quark Yukawa coupling~\cite{radewsb}.  In this
way, the origin of the electroweak symmetry breaking scale is
intimately tied to the mechanism of supersymmetry breaking.  That is,
supersymmetry provides an explanation for the stability of the
hierarchy of scales, provided that supersymmetry-breaking masses in
the low-energy effective electroweak theory are of $\mathcal{O}(1~{\rm
TeV})$ or less~\cite{natural,susynatural}.  One notable
feature of the simplest weak-scale supersymmetric models is the
successful unification of the electromagnetic, weak and strong gauge
interactions, strongly supported by the prediction of $\sin^2\theta_W$
at low energy scales with an accuracy at the percent
level~\cite{susyguts}.
%~\cite{IbanezRoss,susygut}.
%The significance of this prediction is not easily matched by other
%approaches.  For example, in strongly-coupled electroweak symmetry breaking
%models, unification of couplings is not addressed {\it per se}, whereas in  
%extra-dimensional models it is often achieved by introducing new 
%structures at intermediate energy scales.
Unless one is willing to regard
the apparent gauge coupling unification as a coincidence, it is
tempting to conclude that electroweak symmetry breaking is indeed
weakly-coupled, and new physics exists at or below a few 
TeV associated with the supersymmetric extension of the
Standard Model.

A program of Higgs physics at future colliders must address a number
of fundamental questions:
\ben
\item
Does the SM Higgs boson (or a Higgs scalar with similar properties)
exist?
\item
How can one prove that a newly discovered scalar is a Higgs boson?
\item
How many physical Higgs states are associated with the scalar sector?
\item
How well can one distinguish the SM Higgs sector from a more
complicated scalar sector, if only one scalar state is discovered?
\item
Is the Higgs sector consistent with the constraints of 
supersymmetry?
\item
How well can one measure the mass, width, quantum numbers 
and couplings strengths of the Higgs boson?
\item
Are there CP-violating phenomena associated with the Higgs sector?
\item
Can one reconstruct the Higgs potential and directly demonstrate the
mechanism of electroweak symmetry breaking?
\een

\noindent
The physics of the Higgs bosons will
be explored by experiments now underway at the upgraded
proton-antiproton Tevatron collider at Fermilab and
in the near future at the Large Hadron Collider (LHC) at CERN.
Once evidence for the existence of new scalar particles is
obtained, a more complete understanding of the scalar dynamics
will require experimentation at a future $e^+e^-$ linear
collider.  
The next generation of high energy $e^+e^-$ linear colliders
is expected to operate
at energies from 300~GeV up to about 1~TeV (JLC, NLC, TESLA),
henceforth referred to as the LC~\cite{lincoll}.
With the expected high luminosities up to 1~ab$^{-1}$,
accumulated within a few years in a clean experimental environment,
these colliders are ideal instruments for reconstructing the
mechanism of electroweak symmetry breaking in a comprehensive and
conclusive form.

A recent comprehensive review of Higgs theory and phenomenology can be
found in \Ref{carenahaber}, and provides an update to many topics
treated in \textit{The Higgs Hunter's Guide}~\cite{hhg}.
In this short review, I shall highlight
some of the most prominent aspects of the theory and phenomenology of Higgs
bosons of the Standard Model and the MSSM.

\section{The Standard Model Higgs Boson}

\subsection{Theory of the SM Higgs boson}

In the Standard Model, the Higgs mass is given by: $\mhsm^2=\half\lambda
v^2$, where $\lambda$ is the Higgs self-coupling parameter.  Since
$\lambda$ is unknown at present, the value of the SM Higgs
mass is not predicted.  However, other theoretical considerations,
discussed in \Sec{sec:intro}, place constraints on the Higgs mass
as exhibited in \fig{trivial}.  In contrast, the Higgs couplings to
fermions [bosons] are predicted by the theory to be
proportional to the corresponding particle masses [squared-masses].
In particular, the SM Higgs boson is a CP-even scalar, and its
couplings to gauge bosons, Higgs bosons and
fermions are given by:
\beqa 
&& g_{h f\bar f}= {m_f\over v}\,,\qquad\qquad
\qquad\qquad
g_{hVV} = {2m_V^2\over v}\,, \qquad\qquad
g_{hhVV} = {2m_V^2\over v^2}
\,,\nonumber\\[5pt]
&& g_{hhh} = \nicefrac{3}{2}\lambda v ={3\mhsm^2\over v}\,,\,\,\,
\quad\qquad
g_{hhhh} = \nicefrac{3}{2}\lambda= {3\mhsm^2\over
v^2}\,,\label{hsmcouplings}
\eeqa
where $h\equiv\hsm$, $V=W$ or $Z$ and $v=2m_W/g=246$~GeV.
In Higgs production and decay processes, the dominant mechanisms involve
the coupling of the Higgs boson to the $W^\pm$, $Z$ and/or
the third generation quarks and leptons.
Note that a $\hsm gg$ coupling ($g$=gluon)
is induced by virtue of a one-loop graph
in which the Higgs boson couples to a virtual $t\bar t$ pair.
Likewise, a $\hsm\gamma\gamma$ coupling is generated, although in this
case the one-loop graph in which the Higgs boson couples to
a virtual $W^+W^-$ pair is the dominant contribution.    

\begin{figure}[b!]
\begin{center}
\resizebox{\textwidth}{!}{
\includegraphics*{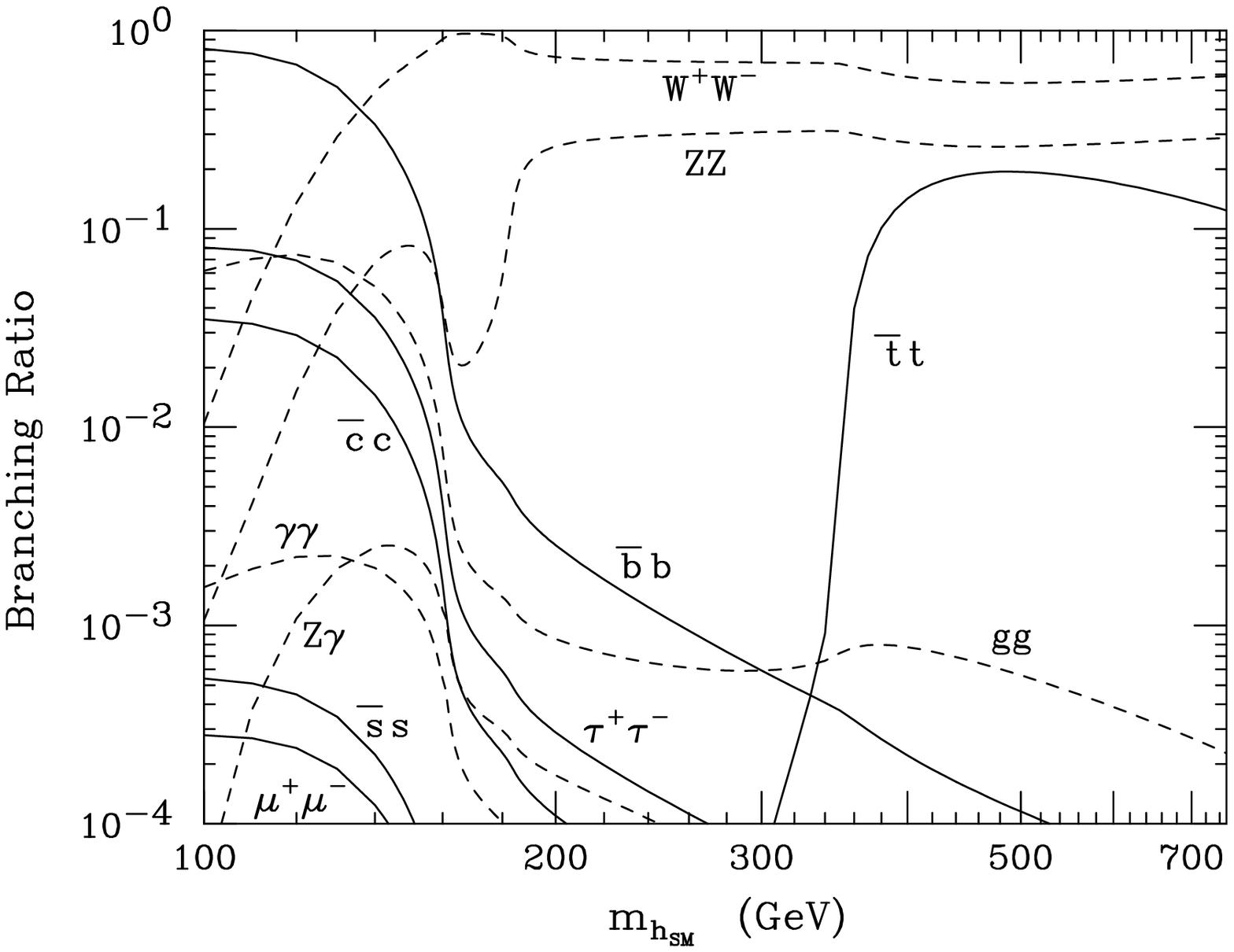}
\hspace*{3mm}
\includegraphics*{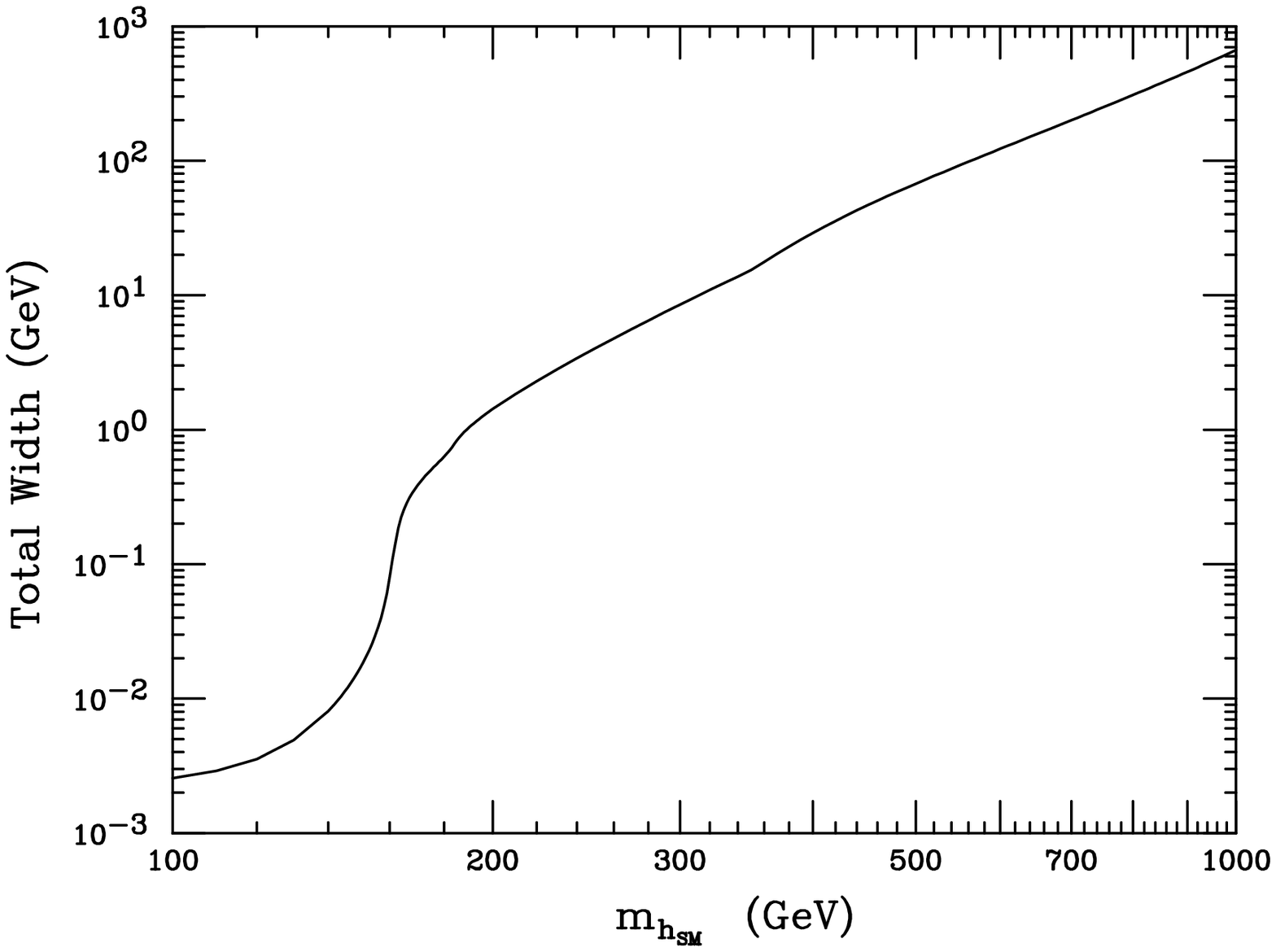}
}
\end{center}
\capt{\label{fg:hwidth} (a) Branching ratios of the SM Higgs
boson as a function of Higgs mass.  Two-boson [fermion-antifermion]
final states are exhibited by solid [dashed] lines.
(b) The total width of the SM Higgs boson is shown as a function of its mass.} 
\end{figure}

The branching ratios for the main decay modes of a SM
Higgs boson are shown as a function of Higgs boson mass in 
\fig{fg:hwidth}(a), based on the results obtained using
the \texttt{HDECAY} program~\cite{hdecay}.
The total Higgs width is obtained by summing all the Higgs partial
widths and is displayed as a function of Higgs mass in \fig{fg:hwidth}(b).

\subsection{Phenomenology of the SM Higgs boson at future colliders}

The Higgs boson will be discovered first at a hadron collider.  At the
Tevatron, the most promising SM Higgs
discovery mechanism for $\mhsm\lsim 135$~GeV consists
of $q\bar q$ annihilation into a virtual $V^*$ ($V=W$ or $Z$), where
$V^*\to V\hsm$ followed by a leptonic decay of the $V$ and
$\hsm\to b\bar b$~\cite{Stange:1994ya}.  These processes lead to three
main final states, $\ell\nu\bb$, $\nn\bb$ and $\lpm\bb$,
that exhibit distinctive signatures on which
the experiments can trigger (high $p_T$ leptons and/or missing $E_T$).
The backgrounds are manageable and are typically dominated by
vector-boson pair production, $t\bar t$ production and QCD dijet
production.  For larger Higgs masses ($\mhsm\gsim 135$~GeV) it is possible to
exploit the distinct signatures present when the Higgs boson decay
branching ratio to $WW^{(*)}$ becomes appreciable.  In this case,
there are final states with $WW$ from the gluon-fusion production of
a single Higgs boson, and $WWW$ and $ZWW$ arising from associated
vector boson--Higgs boson production.
Three search channels were identified in \Ref{tevreport} as
potentially sensitive at these high Higgs masses: like-sign dilepton
plus jets ($\lsdilep$) events, high-$p_T$ lepton pairs plus missing
$E_T$ ($\lpm\nn$), and trilepton ($\trilep$) events.  Of these, the
first two were found to be most sensitive~\cite{turcot}.  The strong
angular correlations of the final state leptons resulting from
$WW^*$ is one of the crucial ingredients for these
discovery channels~\cite{turcot,nelson,dreiner}.

\begin{figure}[t!]
\begin{center}
\resizebox{\textwidth}{3.5in}{
\scalebox{1.40}[2.02]{
\includegraphics*{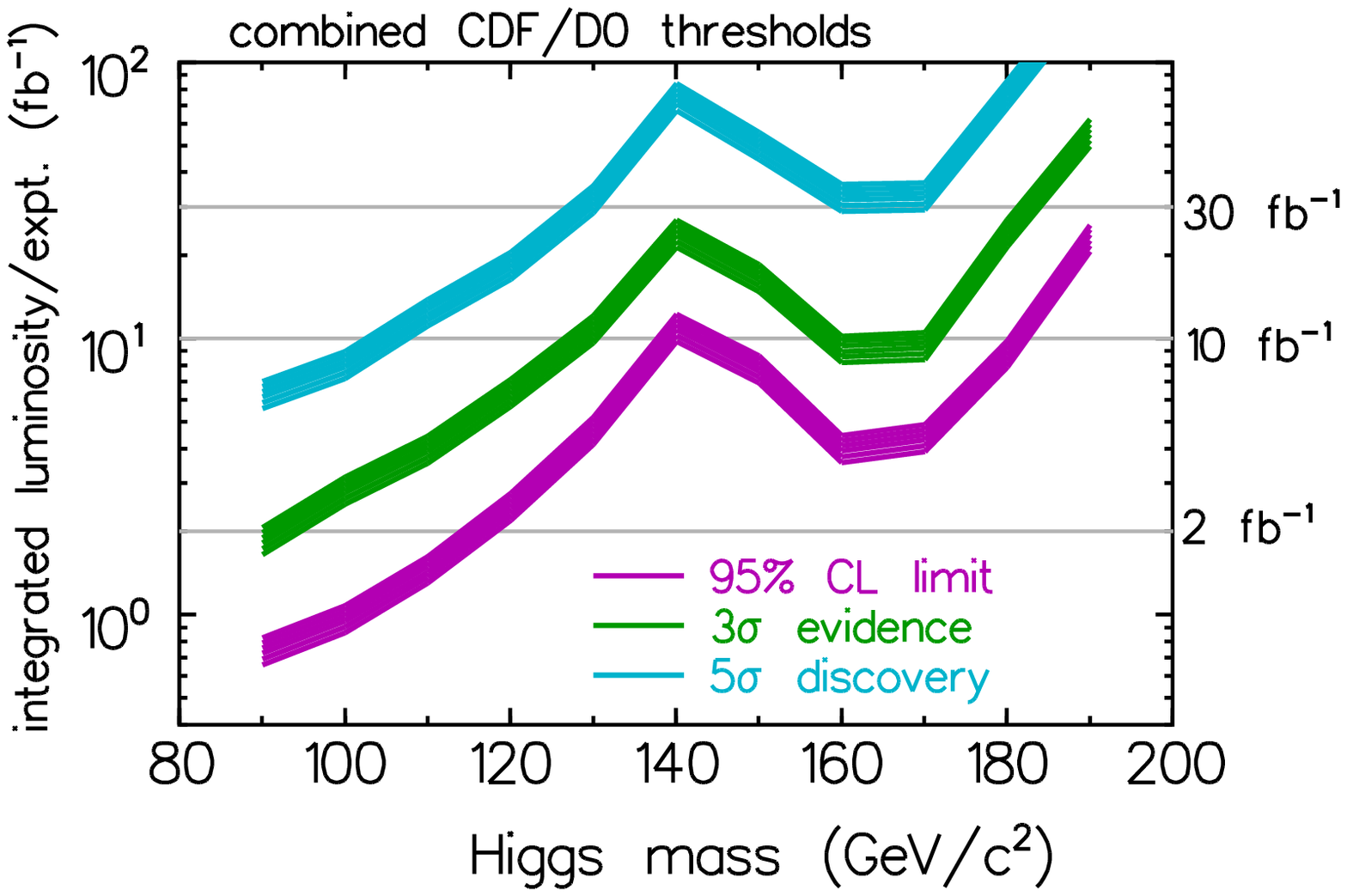}}
\scalebox{1.12}[1.12]{
\includegraphics*{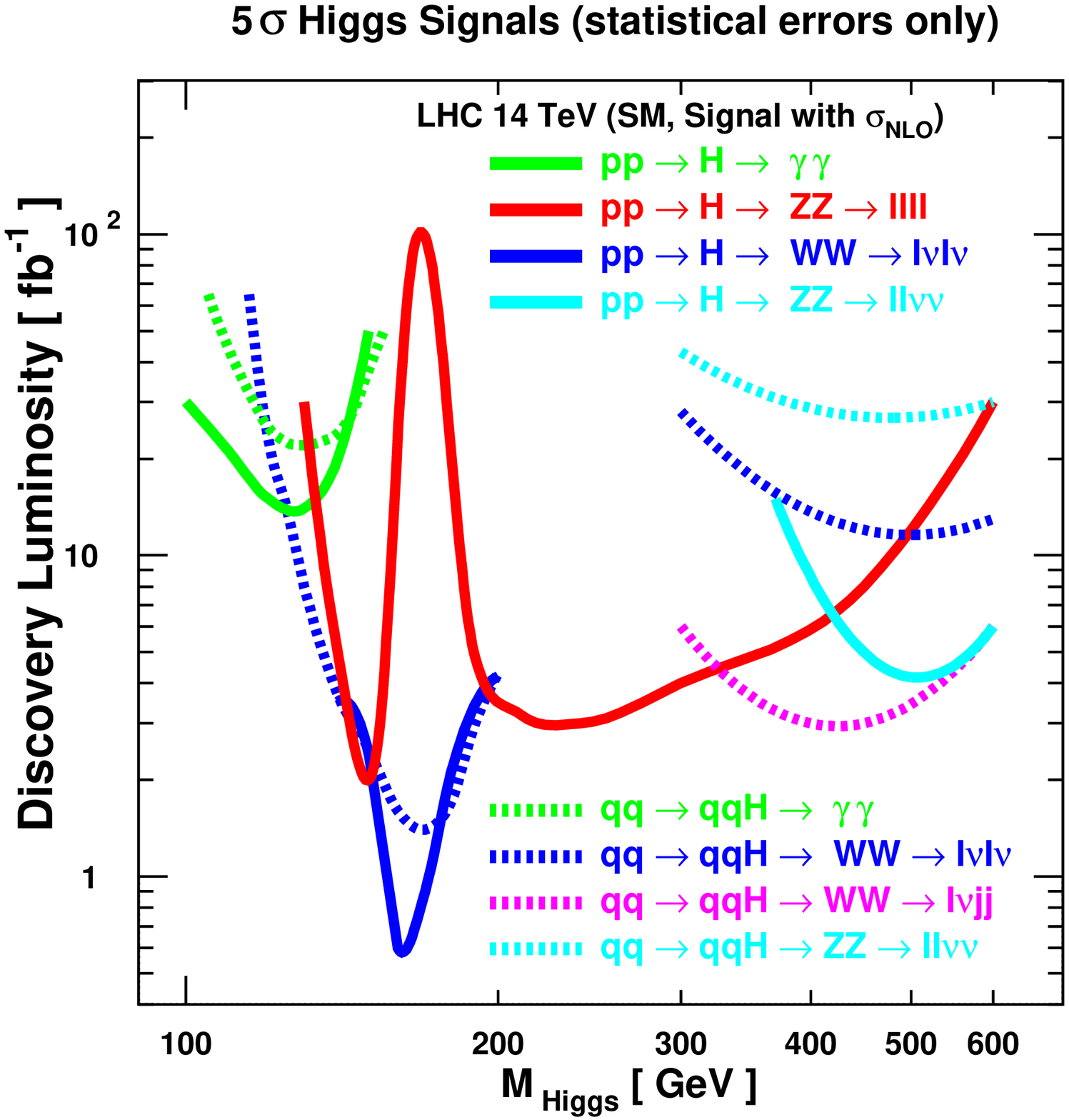}}
}
\end{center}
\capt{\label{fig:smhiggsathadron} (a)~The
integrated luminosity required per Tevatron experiment, to
either exclude a SM Higgs boson at 95\% CL or observe it at the
$3\sigma$ or $5\sigma$ level, as a function of the Higgs
mass~\protect\cite{tevreport}.  (b)~Expected $5\sigma$ discovery 
luminosity requirements for the SM Higgs boson 
at the LHC for one experiment, based on
a study performed with CMS fast detector simulation, assuming
statistical errors only~\protect\cite{Higgs_lumi}.  The $gg$ and
$W^+W^-$ fusion processes are indicated respectively 
by the solid and dotted lines.}
\end{figure}

The integrated luminosity required per Tevatron experiment,
as a function of Higgs mass
to either exclude the SM Higgs boson at 95\% CL or discover it at the
$3\sigma$ or $5\sigma$ level of significance, 
%(see \Ref{tevreport} for details), 
is shown in Fig.~\ref{fig:smhiggsathadron}(a).
These results are based on the combined statistical power of {\it both}
the CDF and D\O\ experiments.
The bands provide an indication of the range of uncertainty in the
$b$-tagging efficiency, $b\bar b$ mass resolution and background
uncertainties.  It is expected that the Tevatron will reach an integrated  
luminosity of $2~{\rm fb}^{-1}$ during its Run 2a phase.  
This will not be sufficient to extend the Higgs search much beyond the
present LEP limits. 
There are plans to further increase the Tevatron luminosity, 
with a possibility of the total integrated
luminosity reaching $6.5$--$11~{\rm fb}^{-1}$ by the end of
2008~\cite{witherell}.  This would provide some
opportunities for the Tevatron to discover or see significant hints of
Higgs boson production.

Soon after the LHC begins operation in 2007, the main Higgs
search efforts will shift to CERN.  A number of different Higgs 
production and decay channels can be studied at the LHC.
The preferred channels for $\mhsm\lsim 200$~GeV are
\beqa
&& gg\to\hsm\to\gamma\gamma\,, \nonumber \\
&& gg\to\hsm\to VV^{(*)}\,, \nonumber \\
&& qq\to qqV^{(*)} V^{(*)}\to qq\hsm,\quad \hsm\to\gamma\gamma,\,
\tau^+\tau^-, \,VV^{(*)}\,, \nonumber \\
&& gg, q\bar q\to t\bar t\hsm, \quad \hsm\to b\bar b,
\,\gamma\gamma, \,WW^{(*)}\,,\nonumber
\eeqa
where $V=W$ or $Z$.
The gluon-gluon fusion mechanism is the dominant Higgs production
mechanism at the LHC.
A recent NNLO computation of $gg\to\hsm$ production demonstrates the
prediction for this Higgs cross-section is under 
theoretical control~\cite{kilgore}.  One also has
appreciable Higgs production via $VV$ electroweak gauge boson fusion,
which can be separated from the
gluon fusion process by employing a forward jet tag
and central jet vetoing techniques.  Finally,
the cross-section for $t\bar t\hsm$ production~\cite{pptth} 
can be significant for
$\mhsm\lsim 200$~GeV, although this cross-section falls faster with Higgs
mass as compared to the gluon and gauge boson fusion mechanisms.
Note that for $2\mw\lsim\mhsm\lsim 2\mz$, the Higgs branching ratio
to $ZZ^*$ is quite suppressed with respect to $WW$ (since one of the $Z$
bosons is off-shell).  Hence, in this mass window, $\hsm\to
W^+W^-\to \ell^+\nu\ell^-\bar\nu$ is the main Higgs discovery 
channel~\cite{dreiner}, as exhibited 
in \fig{fig:smhiggsathadron}(b)~\cite{Higgs_lumi}. 

The measurements of Higgs decay branching ratios at the LHC
can be used to infer the values of the Higgs couplings and
provide an important first step in clarifying the nature of the Higgs
boson~\cite{snow96,zeppenfeld}.  These can be extracted from a variety of
Higgs signals that are observable over a limited range of Higgs
masses.  For example, for $\mhsm\lsim 150$~GeV, the expected
accuracies of Higgs couplings to $W^+W^-$, $\gamma\gamma$,
$\tau^+\tau^-$ and $gg$ can be determined to an accuracy in the range
of 5--15\%~\cite{zeppenfeld}.  
These results are obtained under the assumption that the
partial Higgs widths to $W^+W^-$ and $ZZ$ are fixed by electroweak
gauge invariance, and the ratio of the partial Higgs widths to $b\bar
b$ and $\tau^+\tau^-$ are fixed by the universality of Higgs couplings
to down-type fermions.  One can then extract the total Higgs width
under the assumption that all other unobserved modes, in the Standard
Model and beyond, possess small branching ratios of order 1\%.  The
resulting accuracy anticipated is in the range of 10--25\%, depending
on the Higgs mass.

To significantly improve the precision of Higgs measurements, one
must employ the LC~\cite{Tesla-TDR,charting,vankooten}.
The main production mechanisms of the SM Higgs boson at the LC
are the Higgs-strahlung process~\cite{thacker,strahlung},
$e^+e^- \to Z\hsm$,
and the $WW$ fusion process~\cite{fusion} 
$e^+ e^- \to \bar\nu_e \nu_e W^* W^*\to \bar{\nu}_e \nu _e \hsm$.
With an accumulated luminosity of 500 fb$^{-1}$, about $10^5$ Higgs
bosons can be produced by Higgs-strahlung
in the theoretically
preferred intermediate mass range below 200 GeV. As $\sqrt{s}$ is increased,
the cross-section for the Higgs-strahlung process decreases as $s^{-1}$ and
is dominant at low energies, while the cross-section for the $WW$
fusion process grows as $\ln(s/\mhsm^2)$ 
and dominates at high energies.   
For $\sqrt{s} =350$ and 500 GeV and an integrated
luminosity of 500~fb$^{-1}$, this ensures the
observation of the SM Higgs
boson up to the production kinematical limit independently of its
decay \cite{Tesla-TDR}.
Finally, the process $e^+e^- \rightarrow t \bar t \hsm$~\cite{ttH}
yields a distinctive signature consisting
of two $W$ bosons and four $b$-quark jets, and can be observed at the
LC given sufficient energy and luminosity if the Higgs mass is not too
large ($\mhsm\lsim 200$~GeV at $\sqrt{s}=800$~GeV).

The phenomenological profile of the Higgs boson can be determined by
precision measurements~\cite{snow96}.  For example, 
the Higgs width
can be inferred in a model-independent way, with an accuracy in the 
range of 5--10\% (for $\mhsm\lsim 150$~GeV), by combining
the partial width to $W^+W^-$, accessible in the vector boson fusion
process, with the $WW^*$ decay branching ratio~\cite{desch}.
The spin and parity of the Higgs boson can be
determined unambiguously from the steep onset of the excitation curve in
Higgs-strahlung near the threshold
%(see \fig{fig:LChiggsproperties}(a)~\cite{hspin})
and the angular correlations in this process~\cite{angcorr}.
By measuring final state angular distributions and various 
angular and polarization asymmetries,
one can check whether the Higgs boson is a state of definite CP,
or whether it exhibits CP-violating
behavior in its production and/or decays~\cite{cpasymmetries}.

\begin{table}[b!]
        \begin{center} \begin{tabular}{|c||c|c|}
\hline
        Higgs coupling  & $\delta {\rm BR}/\rm{BR}$ & $\delta g/g$ \\
\hline\hline
        $hWW$     &      5.1\%     &  1.2\% \\
        $hZZ$     &        ---     &  1.2\% \\ \hline
        $htt$     &      ---     &  2.2\% \\
        $hbb$     &      2.4\%     &  2.1\% \\
        $hcc$     &      8.3\%     & 3.1\% \\
        $h \tau\tau$ &      5.0\%     &  3.2\% \\
        $h \mu\mu$   &      $\sim 30\%$  &  $\sim 15\%$
\\ \hline
        $h gg$       &      5.5\% & \\
        $h \gamma\gamma$ &   16\% & \\ 
        $hhh$    &   --- & $\sim 20\%$ \\ \hline
        \end{tabular} \end{center}
\capt{ \label{tab:BRmeas} Expected fractional uncertainties
for measurements of Higgs 
branching ratios [BR($h\to X\overline X$)]
and couplings [$g_{hXX}$], for various choices of final state
$X\overline X$,
assuming $m_h=120$~GeV at the LC.
In all but four cases, the results shown are based on  500~fb$^{-1}$ of data
at $\sqrt{s}=500$~GeV~\protect\cite{BattagliaDesch}. 
The results for $h\gamma\gamma$~\protect\cite{boosphotons},
$h t\bar t$~\protect\cite{BattagliaDesch}, 
$h \mu\mu$~\protect\cite{mumulc} and $hhh$~\protect\cite{bby}
assume 1~ab$^{-1}$ of data
at $\sqrt{s}=500$~GeV (for $\gamma\gamma$ and $hh$) and $\sqrt{s}=800$~GeV
(for $tt$ and $\mu\mu$), respectively.} 
\end{table}

Higgs decay branching ratios can be well measured for
$\mhsm\lsim 150$~GeV~\cite{boosphotons,BattagliaDesch,mumulc,Brau}.  When
such measurements are combined with measurements of Higgs production
cross-sections, the absolute values of the Higgs couplings to the
$W^\pm$ and $Z$ gauge bosons and the Yukawa couplings to leptons and
quarks can be determined to a few percent in a model-independent way.
In addition, the Higgs-top quark Yukawa coupling can be inferred from
the cross-section for Higgs emission off $t\bar t$
pairs~\cite{tth}.  As an example, Table~\ref{tab:BRmeas}
exhibits the anticipated fractional uncertainties in the measurements
of Higgs branching ratios at the LC for $\mhsm=120$~GeV.  Using this
data, a program {\texttt{HFITTER}} was developed in
\Ref{BattagliaDesch} to perform a Standard Model global fit based on
the measurements of the $Z\hsm$, $\nu\bar\nu\hsm$ and $t\bar t\hsm$
cross-sections and the
Higgs branching ratios listed in Table~\ref{tab:BRmeas}.  The output
of the program is a set of Higgs couplings along with their fractional
uncertainties.
These results should be considered representative of what can
eventually be achieved at the LC, after a more complete analysis
incorporating radiative corrections has been performed. 

Finally, the measurement of the Higgs self-couplings is a very
ambitious task that requires the highest luminosities possible 
at the LC~\cite{bby}. 
%which possess unique capabilities for addressing this question. 
The trilinear Higgs self-coupling can be measured in double
Higgs-strahlung, in which a virtual Higgs boson splits into two real
Higgs particles in the final state~\cite{trilinear}.  
The result of a simulation based on 1~ab$^{-1}$ of data~\cite{bby} 
is listed in Table~\ref{tab:BRmeas}. 
Such a measurement is a prerequisite for determining the form of the
Higgs potential that is responsible for spontaneous electroweak
symmetry breaking generated by the scalar sector dynamics.

\section{Higgs Bosons of the MSSM Supersymmetry}
\label{sec:3}

\subsection{Theory of the MSSM Higgs sector}
\label{sec:31}

The simplest realistic supersymmetric model of the fundamental
particles is a minimal supersymmetric extension of the
Standard Model (MSSM)~\cite{susyreview}, which employs the minimal particle
spectrum and soft-supersymmetry-breaking terms (to parameterize the
unknown fundamental mechanism of supersymmetry breaking~\cite{susybreak}).
In constructing the MSSM, both
hypercharge $Y=-1$ and $Y=+1$ complex Higgs doublets are
required in order to obtain
an anomaly-free supersymmetric extension of the Standard Model.
Thus, the MSSM
contains the particle spectrum of a two-Higgs-doublet extension of the
Standard Model and the corresponding supersymmetric partners.

The two-doublet Higgs sector \cite{thdmsusy} contains eight scalar
degrees of freedom:
one complex $Y=-1$ doublet, {\boldmath $\Phi_d$}$=(\Phi_d^0,\Phi_d^-)$
and one complex $Y=+1$ doublet, {\boldmath
$\Phi_u$}$=(\Phi_u^+,\Phi_u^0)$.  The notation reflects the
form of the MSSM Higgs sector coupling to fermions: $\Phi_d^0$
[$\Phi_u^0$] couples exclusively to down-type [up-type] fermion pairs.
When the Higgs potential is minimized, the neutral 
Higgs fields acquire vacuum expectation values:\footnote{The
phases of the Higgs fields can be chosen such that the vacuum
expectation values are real and positive.  That is, the tree-level MSSM
Higgs sector conserves CP, which implies that the neutral Higgs mass
eigenstates possess definite CP quantum numbers.}
\beq
\langle {\mathbold{\Phi_d}} \rangle={1\over\sqrt{2}} \left(
\begin{array}{c} v_d\\ 0\end{array}\right), \qquad \langle
{\mathbold{\Phi_u}}\rangle=
{1\over\sqrt{2}}\left(\begin{array}{c}0\\ v_u
\end{array}\right)\,,\label{potmin}
\eeq
where $\tanb\equiv v_u/v_d$ and the normalization has been chosen such that
$v^2\equiv v_d^2+v_u^2={4\mw^2/ g^2}=(246~{\rm GeV})^2$.
Spontaneous electroweak symmetry breaking results in
three Goldstone bosons, which
are absorbed and become the longitudinal components of
the $W^\pm$ and $Z$.  The remaining five physical Higgs particles
consist of a charged Higgs pair, $\hpm$, one CP-odd scalar, $\ha$ 
and two CP-even scalars:
\beqa
\hl &=& -(\sqrt{2}\,{\rm Re\,}\Phi_d^0-v_d)\sin\alpha+
(\sqrt{2}\,{\rm Re\,}\Phi_u^0-v_u)\cos\alpha\,,\nonumber\\[4pt]
\hh &=& (\sqrt{2}\,{\rm Re\,}\Phi_d^0-v_d)\cos\alpha+
(\sqrt{2}\,{\rm Re\,}\Phi_u^0-v_u)\sin\alpha\,,
\label{scalareigenstates}
\eeqa
(with $\mhl\leq \mhh$).
The angle $\alpha$ arises when the CP-even Higgs
squared-mass matrix (in the $\Phi_d^0$---$\Phi_u^0$ basis) is
diagonalized to obtain the physical CP-even Higgs states.

The supersymmetric structure of the theory imposes constraints on the
Higgs sector.  For example, the Higgs
self-interactions are not independent parameters; they can be expressed
in terms of the electroweak gauge coupling constants.  As a result,
all Higgs sector parameters at tree-level
are determined by two free parameters, which may be taken to be
$\tanb$ and $\mha$.  One significant consequence of these results
is that there is a tree-level upper bound
to the mass of the light CP-even Higgs boson, $\hl$.  One finds that:
$\mhl\leq\mz |\cos 2\beta|\leq\mz$.
This is in marked contrast to the Standard Model, in which the theory
does not constrain the value of $\mhsm$ at tree-level.  The origin of
this difference is easy to ascertain.  In the \SM, $\mhsm^2=\half\lambda v^2$
is proportional to the Higgs self-coupling $\lambda$, which is a free
parameter.  On the other hand, all Higgs self-coupling
parameters of the MSSM are related to the squares of the electroweak
gauge couplings.

%Note that the Higgs mass inequality [\eq{kx}] is saturated in the limit
%of large $\mha$.  
In the limit of $\mha\gg\mz$, the expressions for the
Higgs masses simplify and one finds:
\beq
\mhl^2 \simeq  \ \mz^2\cos^2 2\beta\,,\qquad
\mhh^2 \simeq  \ \mha^2+\mz^2\sin^2 2\beta\,,\qquad
\mhpm^2 =  \ \mha^2+\mw^2\,.\label{largema}
\label{largema4}
\eeq
In addition, the behavior of the 
quantity $\cosbma$ is noteworthy in this limit.  One can
show that at tree level,
\beq
\cos^2(\beta-\alpha)={\mhl^2(\mz^2-\mhl^2)\over
\mha^2(\mhh^2-\mhl^2)}\simeq {m_Z^4\sin^2 4\beta\over 4\mha^4}\,,
\label{cbmasq}
\eeq
where the last result on the right-hand side above corresponds to the
limit of large $\mha$.
Two consequences of these results are immediately apparent.
First, $\mha\simeq\mhh
\simeq\mhpm$, up to corrections of ${\cal O}(\mz^2/\mha)$.  Second,
$\cos(\beta-\alpha)=0$ up to corrections of ${\cal O}(\mz^2/\mha^2)$.
This limit is known as the {\it decoupling} limit \cite{decoupling}
because when $\mha$ is
large, there exists an effective low-energy theory below the scale
of $\mha$ in which the effective Higgs sector consists only of one
CP-even Higgs boson, $\hl$.  In particular, one can check that 
when $\cosbma=0$, the
tree-level couplings of $\hl$ are precisely those of the 
SM Higgs boson.

%\begin{figure}[t!]
%\begin{center}
%\resizebox{\textwidth}{!}{
%\includegraphics[width=5cm]{cos2bma3.ps}
%\hfill
%\includegraphics[width=5cm]{cos2bma30.ps}
%}
%\end{center}
%\capt{\label{cosgraph} The value of $\cos^2(\beta-\alpha)$
%is shown as a function of
%$\mha$ for two choices of $\tan\beta = 3$ and $\tan\beta = 30$.
%When radiative-corrections are included, one can define an approximate
%loop-corrected angle $\alpha$ as a function of $\mha$, $\tan\beta$ and
%the MSSM parameters.  In the figures above,
%radiative corrections have been incorporated, assuming that
%the supersymmetry-breaking scale is $M_S= 1$~TeV.  In addition,
%two extreme cases for the squark mixing parameters
%are shown. The decoupling effect expected from
%\protect\eq{largema4}, in which $\cos^2(\beta-\alpha)\propto \mz^4/\mha^4$
%for $\mha\gg m_Z$,
%continues to hold when radiative corrections are included.}
%\end{figure}

The phenomenology of the Higgs sector depends in detail on
the various couplings of the Higgs bosons to gauge bosons, Higgs
bosons and fermions.  The couplings of the Higgs bosons
to $W$ and $Z$ pairs typically depend on
the angles $\alpha$ and $\beta$.  The properties of the three-point and
four-point Higgs boson--vector boson couplings are conveniently
summarized by listing the various couplings that are proportional
to either $\sin(\beta-\alpha)$ or $\cos(\beta-\alpha)$, and those
couplings that are independent of $\alpha$ and $\beta$~\cite{hhg}:
\begin{displaymath}
\renewcommand{\arraycolsep}{1cm}
\let\us=\underline
\begin{array}{lll}
\us{\cos(\beta-\alpha)}&  \us{\sin(\beta-\alpha)} &
\us{\hbox{\rm{angle-independent}}} \\ [3pt]
\noalign{\vskip3pt}
       \hh W^+W^-&    \hl W^+W^- & \qquad\longdash \\  
       \hh ZZ&            \hl ZZ  & \qquad\longdash \\
       Z\ha\hl&          Z\ha\hh  & ZH^+H^-\,,\,\,\gamma H^+H^-\\
       W^\pm H^\mp\hl&  W^\pm H^\mp\hh & W^\pm H^\mp\ha \\
       ZW^\pm H^\mp\hl&  ZW^\pm H^\mp\hh & ZW^\pm H^\mp\ha \\
    \gamma W^\pm H^\mp\hl&  \gamma W^\pm H^\mp\hh & \gamma W^\pm H^\mp\ha \\
    \quad\longdash   & \quad\longdash& VV\phi\phi\,,\,VV\ha\ha\,,\,VV H^+H^-
\end{array}
\end{displaymath}
where $\phi=\hl$ or $\hh$ and $VV=W^+W^-$, $ZZ$, $Z\gamma$ or
$\gamma\gamma$.
Note that {\it all} vertices
in the theory that contain at least
one vector boson and {\it exactly one} non-minimal Higgs boson state
($\hh$, $\ha$ or $\hpm$) are proportional to $\cos(\beta-\alpha)$.

In the MSSM, the tree-level Higgs couplings to fermions obey the
following
property: $\Phi_d^0$ couples exclusively to down-type fermion pairs and
$\Phi_u^0$ couples exclusively to up-type fermion pairs. This pattern
of Higgs-fermion couplings defines the
Type-II two-Higgs-doublet model \cite{wise,hhg}.
The gauge-invariant Type-II Yukawa interactions (using 3rd family
notation) are given by:
\begin{equation} \label{typetwo}
-{\cal L}_{\rm Yukawa}= h_t\left[\bar t_R t_L \Phi^0_u-\bar t_R b_L
\Phi^+_u\right] + h_b\left[\bar b_R b_L \Phi^0_d-\bar b_R t_L
\Phi^-_d\right] + {\rm h.c.}\,,
\end{equation}
where $q_{R,L}\equiv\half(1\pm\gamma_5)q$. 
Inserting \eq{potmin} into \eq{typetwo}
yields a relation between the quark masses and the Yukawa couplings:
\beq
h_b = {\sqrt{2}\,m_b\over v_d}={\sqrt{2}\, m_b\over
v\cos\beta}\,,\qquad\qquad
h_t = {\sqrt{2}\,m_t\over v_u}={\sqrt{2}\, m_t\over v\sin\beta}\,.
\label{hdef}
\eeq
Similarly, one can define the Yukawa coupling of the Higgs boson to
$\tau$-leptons (the $\tau$ is a down-type fermion).  
The $\hl f\bar f$ couplings relative to the Standard Model
value, $m_f/v$, are then given by
\begin{eqaligntwo}
\hl b\bar b \;\;\; ({\rm or}~ \hl \tau^+ \tau^-):&~~~ -
{\sin\alpha\over\cos\beta}=\sin(\beta-\alpha)
-\tan\beta\cos(\beta-\alpha)\,, \label{qqcouplingsa}\\[3pt]
\hl t\bar t:&~~~ \phm{\cos\alpha\over\sin\beta}=\sin(\beta-\alpha)
+\cot\beta\cos(\beta-\alpha)\,. \label{qqcouplingsb}
\end{eqaligntwo}
As previously noted, $\cos(\beta-\alpha)={\cal O}(\mz^2/\mha^2)$
in the decoupling limit where $\mha\gg\mz$. As a result,
the $\hl$ couplings to
Standard Model particles approach values corresponding precisely
to the couplings of the SM Higgs boson.
There is a significant
region of MSSM Higgs sector parameter space in which the decoupling
limit applies, because
$\cos(\beta-\alpha)$ approaches zero quite rapidly once
$\mha$ is larger than about 200~GeV.
%as shown in \fig{cosgraph}.
As a result, over a significant region of the MSSM parameter space, the
search for the lightest CP-even Higgs boson
of the MSSM is equivalent to the search for the SM
Higgs boson.  

The tree-level analysis of Higgs masses and couplings described above can be
significantly altered once radiative corrections are included.
The dominant effects arise from
loops involving the third generation quarks and squarks
and are proportional to the corresponding Yukawa couplings.
For example, consider the tree-level upper bound on the lightest
CP-even Higgs mass,
$\mhl\leq m_Z$, a result already ruled out by LEP data. This 
inequality receives quantum corrections primarily from an incomplete
cancellation of top quark and top squark loops~\cite{hhprl}
(this cancellation would have been exact if supersymmetry were
unbroken).  Radiative corrections can also generate CP-violating
effects in the Higgs sector due to CP-violating  supersymmetric
parameters, which enter in the loop computations~\cite{cpsusy}.
Observable consequences include Higgs scalar eigenstates
of mixed CP quantum numbers and CP-violating Higgs-fermion couplings.
However, for simplicity, such effects are assumed to be small 
and are neglected in the following discussion.

The qualitative behavior of the radiative corrections can be most easily
seen in the large top squark mass limit, where the
splitting of the two diagonal entries and the off-diagonal entry
of the top-squark squared-mass matrix are both small in comparison to
the average of the two top-squark squared-masses,
$M_S^2\equiv\half(\mstopa^2+\mstopb^2)$.
In this case, the upper bound on the lightest CP-even Higgs
mass is approximately given by
\beq \label{deltamh}
\mhl^2\lsim \mz^2+{3g^2\mt^4\over
8\pi^2\mw^2}\left[\ln\left({M_S^2\over\mt^2}\right)+{X_t^2\over M_S^2}
\left(1-{X_t^2\over 12M_S^2}\right)\right]\,.
\eeq

More complete treatments of the radiative 
corrections~\footnote{Detailed analytic approximations to the
radiatively-corrected Higgs masses can be found in
\Ref{susyhiggsloop}.  A recent review of the status of the most
complete Higgs mass computations has been given in \Ref{susyhiggsprecise}.} 
show that
\eq{deltamh} somewhat overestimates the true upper bound of $\mhl$.
Nevertheless, \eq{deltamh} correctly reflects some noteworthy features
of the more precise result.
First, the increase of the
light CP-even Higgs mass bound beyond $\mz$ can be significant.  This is
a consequence of the $m_t^4$ enhancement of the one-loop radiative
corrections.
Second, the dependence of the light Higgs mass on the top-squark mixing
parameter $X_t$ implies that (for a given value of
$M_S$) the upper bound of the light Higgs mass
initially increases with $X_t$ and reaches its {\it maximal} value for
$X_t\simeq \sqrt{6}M_S$.  This latter is referred to as the {\it maximal
mixing} case (whereas $X_t=0$ corresponds to the {\it
minimal mixing} case).
Third, note the logarithmic sensitivity to the top-squark masses.
Naturalness arguments imply
that the supersymmetric particle masses should not be larger than
a few TeV.  Still, the precise upper bound on the light Higgs mass
depends on the specific choice for the upper limit of the
top-squark masses.

At fixed $\tan\beta$, the maximal value of $\mhl$ is
reached for $\mha\gg\mz$. For large $\mha$, the maximal value of the
lightest CP-even Higgs mass ($\mhmax$)
is realized at large $\tanb$ in
the case of maximal mixing. 
Allowing for the uncertainty in the measured value of
$\mt$ and the uncertainty inherent in the theoretical analysis,
one finds for $M_S\lsim 2$~TeV that~\cite{susyhiggsloop} 
\beqa
\label{mhmaxvalue} \mhmax&\simeq & 122~{\rm GeV}, \quad
\mbox{if top-squark mixing is minimal,} \nonumber \\[3pt]
\mhmax&\simeq & 135~{\rm GeV}, \quad
\mbox{if top-squark mixing is maximal.}
\eeqa
In practice, parameters leading to maximal
mixing are not expected in typical models of
supersymmetry breaking. Thus, in general, the  upper bound
on the lightest Higgs boson mass is expected to be
somewhere between the two extreme limits quoted above.

Radiative corrections can also have a significant impact on the
pattern of Higgs couplings, particularly at large values of $\tanb$.
The leading contributions to the radiatively-corrected Higgs
couplings arise in two ways.  First, to a good approximation
the dominant Higgs propagator corrections 
can be absorbed into an effective 
(``radiatively-corrected'') mixing angle $\alpha$~\cite{hffsusyprop}.
In this approximation, $\cosbma$ is given in terms of the
radiatively-corrected CP-even Higgs squared-mass matrix elements
$\mathcal{M}^2_{ij}$ as follows
\beq \label{eq:cosbma}
\cosbma={(\mathcal{M}_{11}^2-\mathcal{M}_{22}^2)\sin 2\beta
-2\mathcal{M}_{12}^2\cos 2\beta\over
2(m_H^2-m_h^2)\sin(\beta-\alpha)} \,,
\eeq
Defining $\mathcal{M}^2\equiv \mathcal{M}^2_0+\delta \mathcal{M}^2$,
where $\mathcal{M}^2_0$ denotes the tree-level squared-mass matrix,
and noting that $\delta\mathcal{M}^2_{ij}\sim {\mathcal O}(m_Z^2)$
and $m_H^2-m_h^2=m_A^2+\mathcal{O}(m_Z^2)$, one obtains 
for $m_A\gg m_Z$
\beq \label{cbmashift}
\cos(\beta-\alpha)=c\left[{m_Z^2\sin 4\beta\over
2m_A^2}+\mathcal{O}\left(m_Z^4\over m_A^4\right)\right]\,,
\eeq
where
\beq \label{cdef}
c\equiv 1+{{\delta\mathcal{M}}_{11}^2-{\delta\mathcal{M}}_{22}^2\over
2m_Z^2\cos 2\beta}-{{\delta\mathcal{M}}_{12}^2\over m_Z^2\sin
2\beta}\,.
\eeq
Using tree-level Higgs couplings with $\alpha$ replaced by its 
effective one-loop value provides a useful first approximation to 
the radiatively-corrected Higgs couplings.

%For Higgs couplings to vector bosons, the dominant corrections
%arise from corrections to $\cos(\beta-\alpha)$.  
For Higgs couplings to fermions, in addition to the radiatively-corrected
value of $\cos(\beta-\alpha)$,
one must also consider Yukawa vertex corrections.  When these
radiative corrections are included, all possible dimension-four
Higgs-fermion couplings are generated.  In particular, the effects of
higher dimension operators can be ignored if $M_S\gg\mz$, which we
henceforth assume.
These results can be summarized by an effective
Lagrangian that describes the coupling of
the neutral Higgs bosons to the third generation
quarks:
\beq \label{susyyuklag}
        -\mathcal{L}_{\rm eff} = 
        (h_b + \delta h_b) \bar b_R b_L \Phi_d^0
        + (h_t + \delta h_t) \bar t_R t_L \Phi_u^0 
       + \Delta h_t \bar t_R t_L \Phi_d^{0 \ast} 
        + \Delta h_b \bar b_R b_L \Phi_u^{0 \ast}
        + {\rm h.c.}\,,
\eeq
resulting in a modification of the tree-level relation between
$h_q$ and $m_q$ ($q=b$, $t$)~\cite{deltamb}:
\beqa
        m_b &=& \frac{h_b v}{\sqrt{2}} \cos\beta
        \left(1 + \frac{\delta h_b}{h_b}
        + \frac{\Delta h_b \tan\beta}{h_b} \right) 
        \equiv\frac{h_b v}{\sqrt{2}} \cos\beta
        (1 + \Delta_b)\,, \label{byukmassrel} \\[5pt]
        m_t &=& \frac{h_t v}{\sqrt{2}} \sin\beta
        \left(1 + \frac{\delta h_t}{h_t} + \frac{\Delta
        h_t\cot\beta}{h_t} \right)
        \equiv\frac{h_t v}{\sqrt{2}} \sin\beta
        (1 + \Delta_t)\,. \label{tyukmassrel}
\eeqa
The dominant contributions to $\Delta_b$ are $\tan\beta$-enhanced.
In particular, for $\tan\beta\gg 1$,
$\Delta_b\simeq (\Delta h_b/h_b)\tan\beta$; whereas
$\delta h_b/h_b$ provides a small correction to
$\Delta_b$.   In the same limit, $\Delta_t\simeq\delta h_t/h_t$, with
the additional contribution of $(\Delta h_t/h_t)\cot\beta$ providing a
small correction.  Explicitly, 
\beqa
\!\!\!\!\! \Delta_b 
        &\simeq&\left[ 
        \frac{2 \alpha_s}{3 \pi} \mu M_{\tilde g} \,
        I(M^2_{\tilde b_1}, M^2_{\tilde b_2}, M^2_{\tilde g})
        + \frac{h_t^2}{16 \pi^2} \mu A_t \,
        I(M^2_{\tilde t_1}, M^2_{\tilde t_2}, \mu^2)\right]\tan\beta
        \nonumber \\[5pt]
        \Delta_t &\simeq& 
        -\frac{2 \alpha_s}{3 \pi} A_t M_{\tilde g} I(M^2_{\tilde t_1},
        M^2_{\tilde t_2}, M^2_{\tilde g})
        - \frac{h_b^2}{16 \pi^2} \mu^2 I(M^2_{\tilde b_1}, 
        M^2_{\tilde b_2}, \mu^2)\,, \nonumber
\eeqa
where the function $I$ is defined by:
\beq
I(a,b,c) = {ab\ln(a/b)+bc\ln(b/c)+ca\ln(c/a) \over
(a-b)(b-c)(a-c)}\,.
\eeq
Note that $I$ is manifestly positive and $I(a,a,a)=1/(2a)$.

The $\tau$ couplings are obtained 
by replacing $m_b$, $\Delta_b$ and $\delta h_b$
with $m_{\tau}$, $\Delta_{\tau}$ and $\delta h_\tau$, respectively.  
At large $\tan\beta$,
\beq
\Delta_\tau \simeq \left[{\alpha_1 \over 4\pi} M_1\mu
I(M^2_{\tilde\tau_1},
M^2_{\tilde\tau_2},M^2_1) - {\alpha_2 \over 4\pi} M_2\mu \,
I(M^2_{\tilde\nu_\tau},M^2_2,\mu^2)\right]\tan\beta\,,
\eeq
where $\alpha_2\equiv g^2/4\pi$ and $\alpha_1\equiv g^{\prime\,2}/4\pi$
are the electroweak gauge couplings.  In general, one expects that
$|\Delta_\tau|\ll|\Delta_b|$. 

Including the
leading radiative corrections, the $\hl f\bar f$ couplings are given by
\beqa 
\hl b\bar b&:&~~~~ -{m_b\over v}{\sin\alpha \over \cos\beta}
\left[1+{1\over 1+\Delta_b}\left({\delta h_b\over h_b}-
\Delta_b\right)\left( 1 +\cot\alpha \cot\beta \right)\right]
\nonumber \\[6pt]
\hl t\bar t&:&~~~~~~~~ {m_t\over v}{\cos\alpha \over \sin\beta}
\left[1-{1\over 1+\Delta_t}{\Delta h_t\over h_t}
(\cot\beta+ \tan\alpha)\right]\,.
\eeqa
Away from the decoupling limit,
the Higgs couplings to down-type fermions can deviate significantly
from their tree-level values due to enhanced radiative corrections at
large $\tan\beta$ [where $\Delta_b\simeq\mathcal{O}(1)$].
%In particular, because $\Delta_b\propto\tan\beta$, the leading
%one-loop radiative correction to $g_{\hl b\bar b}$ is of
%$\mathcal{O}(\mz^2\tanb/\mha^2)$, which decouples only when
%$\mha^2\gg\mz^2\tanb$ (this behavior was called {\it delayed
%decoupling} in \cite{loganetal}).  
However, in the approach to the
decoupling limit, one can work to first order in $\cos(\beta-\alpha)$ and
obtain 
\beqa
	g_{\hl bb}&\simeq & g_{\hsm bb}\biggl[1+
          (\tan\beta+\cot\beta) \cos(\beta-\alpha)
\left(\cos^2\beta-{1+\delta h_b/h_b\over 1+\Delta_b}
        \right)\biggr]\,, \nonumber \\
	g_{\hl tt}&\simeq & g_{\hsm tt}\left[1+\cos(\beta-\alpha)
\left(\cot\beta-{1\over1+\Delta_t}{\Delta h_t\over h_t}{1\over\sin^2\beta}
\right)\right]\,.
\eeqa
Note that \eq{cbmashift} implies that $(\tanb+\cot\beta)
\cos(\beta-\alpha)\simeq \mathcal{O}(m_Z^2/\mha^2)$,
even if $\tanb$ is very large (or small). 
Thus, at large $\mha$ the deviation of the $\hl b\bar b$ coupling 
from its  SM value vanishes as $\mz^2/\mha^2$ for all values of
$\tanb$. 

Thus, if we keep only the leading $\tanb$-enhanced radiative
corrections, then~\cite{chlm} 
\beqa
   {g^2_{hVV}\over g^2_{\hsm VV}} & \simeq &
1-{c^2 m_Z^4\sin^2 4\beta\over 4m_A^4}\,, \qquad\quad
%\nonumber \\[5pt]      
   {g^2_{htt}\over g^2_{\hsm tt}}  \simeq  1+{c m_Z^2\sin 4\beta
\cot\beta\over m_A^2}\,, \nonumber \\[5pt]
  {g^2_{hbb}\over g^2_{\hsm bb}} & \simeq & 1-{4c m_Z^2\cos 2\beta
\over m_A^2}\left[\sin^2\beta-{\Delta_b\over 1+\Delta_b}\right]\,.
\eeqa    
The approach to decoupling is fastest for the $\hl$ couplings to
vector bosons and slowest for the couplings to down-type quarks.

%When CP-violation effects are non-negligible, we denote the three neutral
%Higgs bosons (of indefinite CP) by $H_1$, $H_2$ and $H_3$, where $H_1$
%is the lightest neutral Higgs boson.  In this case, the decoupling limit is
%governed by the condition $\mhpm\gg m_W$.  In this limit, one can show
%that the properties of $H_1$ are identical to those of the SM Higgs boson.
%That is, $H_1$ is a pure CP-even state, up to corrections of $\mathcal{
%O}(m_W^2/\mhpm^2)$.  Nevertheless, there
%can still be large mixing between $\hh$ and $\ha$ ({\it i.e.},
%$H_2$ and $H_3$ can have large admixtures of CP-even and odd components).
%However, in the decoupling limit,
%$H_2$ and $H_3$ will be nearly mass-degenerate and difficult
%to separate experimentally.

Note that it is possible for $\hl$ to behave like a SM Higgs boson
outside the parameter regime where decoupling has set in.
This phenomenon can arise if the
MSSM parameters (which govern the Higgs mass radiative
corrections) take values such that $c=0$, or equivalently [from \eq{cdef}]:
\begin{equation}
        2\mz^2\sin 2\beta =
        2\, \delta \mathcal{M}^2_{12}
        - \tan 2\beta
        \left(\delta \mathcal{M}^2_{11} - \delta \mathcal{M}^2_{22}\right) 
        \,.
        \label{eq:tanbetadecoup}
\end{equation}
In this case, $\cosbma=0$, due to a cancellation
of the tree-level and one-loop contributions.
%\footnote{The
%two-loop corrections are subdominant, so that the approximation scheme
%is under control.}  
In particular,
\eq{eq:tanbetadecoup} is independent of the value of $m_A$.
Typically,
\eq{eq:tanbetadecoup} yields a solution at large $\tan\beta$.  That
is, by
approximating $\tan 2\beta\simeq -\sin 2\beta \simeq -2/ \tan \beta$,
one can determine
the value of $\tanb$ at which $\cosbma\simeq 0$~\cite{chlm}:
\begin{equation} \label{earlydecoupling}
\tan \beta\simeq {2m_Z^2-
\delta\mathcal{M}_{11}^2+\delta \mathcal{M}_{22}^2\over
\delta\mathcal{M}_{12}^2}\,.
\end{equation}
%Hence, there exists a value of $\tanb$ (which depends on the choice
%of MSSM parameters) where
%$\cos(\beta-\alpha)\simeq 0$ independently of the value of $m_A$.  
If $\mha$ is not much larger than
$\mz$, then $\hl$ is a SM-like Higgs boson outside the decoupling regime.

\subsection{Phenomenology of MSSM Higgs bosons at future colliders}

We have noted that in the decoupling regime, $\hl$ of the MSSM behaves
like the SM Higgs boson.  Thus, for $\mha\gsim 200$~GeV, the Higgs discovery
reach of future colliders is nearly identical to that of the
SM Higgs boson.  In addition, if the heavier Higgs states
are not too heavy, then they can also be directly observed.  It is
convenient to present the discovery contours in the $\mha$--$\tanb$
plane.  At the Tevatron, there is only a small region of the parameter
space in which more than one Higgs boson can be detected.  This is a
region of small $\mha$ and large $\tanb$, where $b\bar b\ha$
production is 
$\tanb$-enhanced~\cite{tevbbbb}.\footnote{In the same region of
parameter space, one of the CP-even Higgs states also has an enhanced
cross-section when produced in association with $b\bar b$.  Finally,
the discovery of the charged Higgs boson 
is possible via $t\to b H^+$ (and $\bar t\to \bar b H^-$) 
decay~\cite{tevreport}
if $\mhpm<m_t-m_b$ [since ${\rm BR}(t\to b H^+)$ is non-negligible
for large (and small) values of $\tanb$].}

\begin{figure}[t!]
\begin{center}
\resizebox{\textwidth}{!}{
\scalebox{1.0}[1.25]{
\includegraphics*{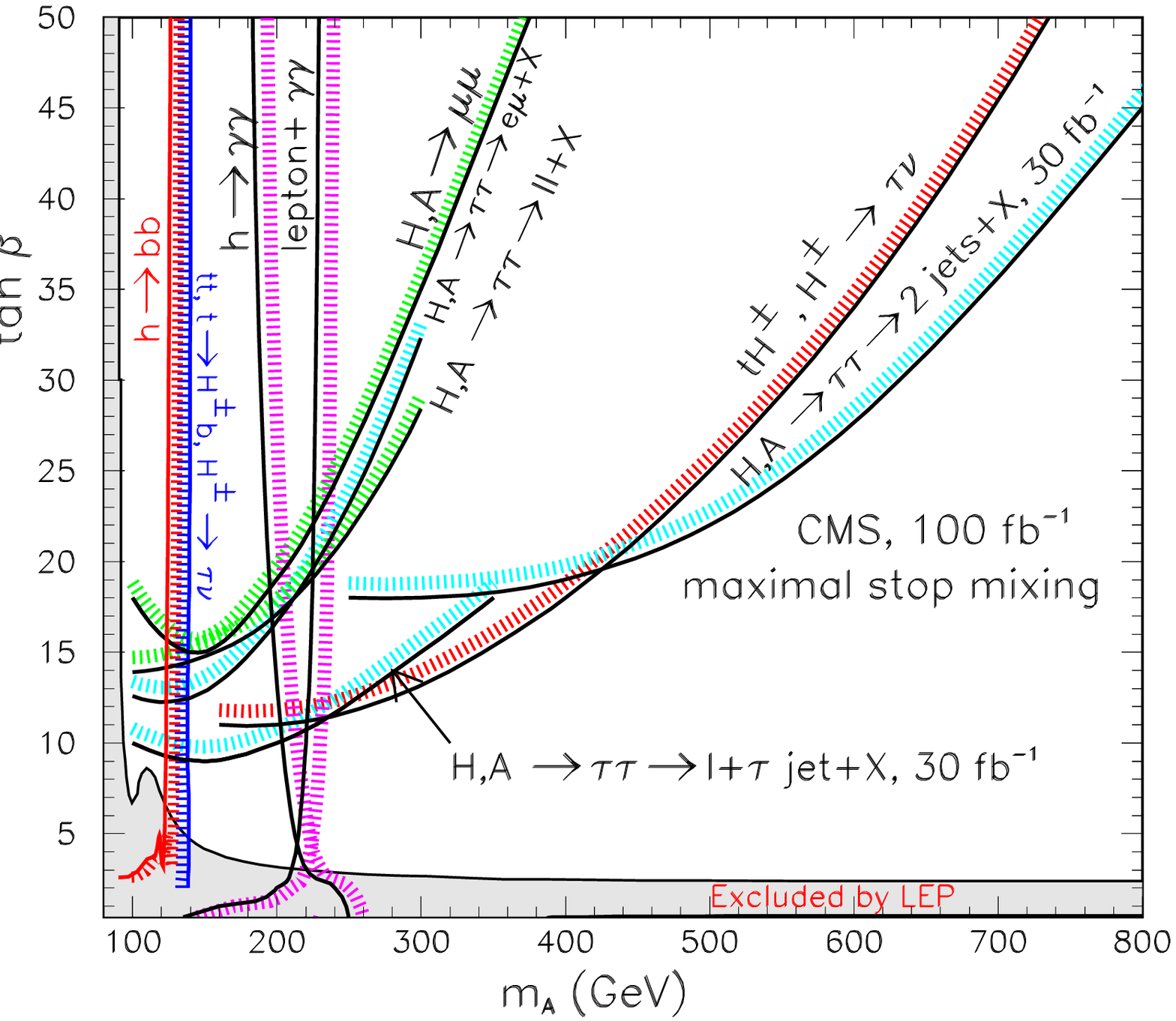}}
\hspace*{3mm}
\includegraphics*{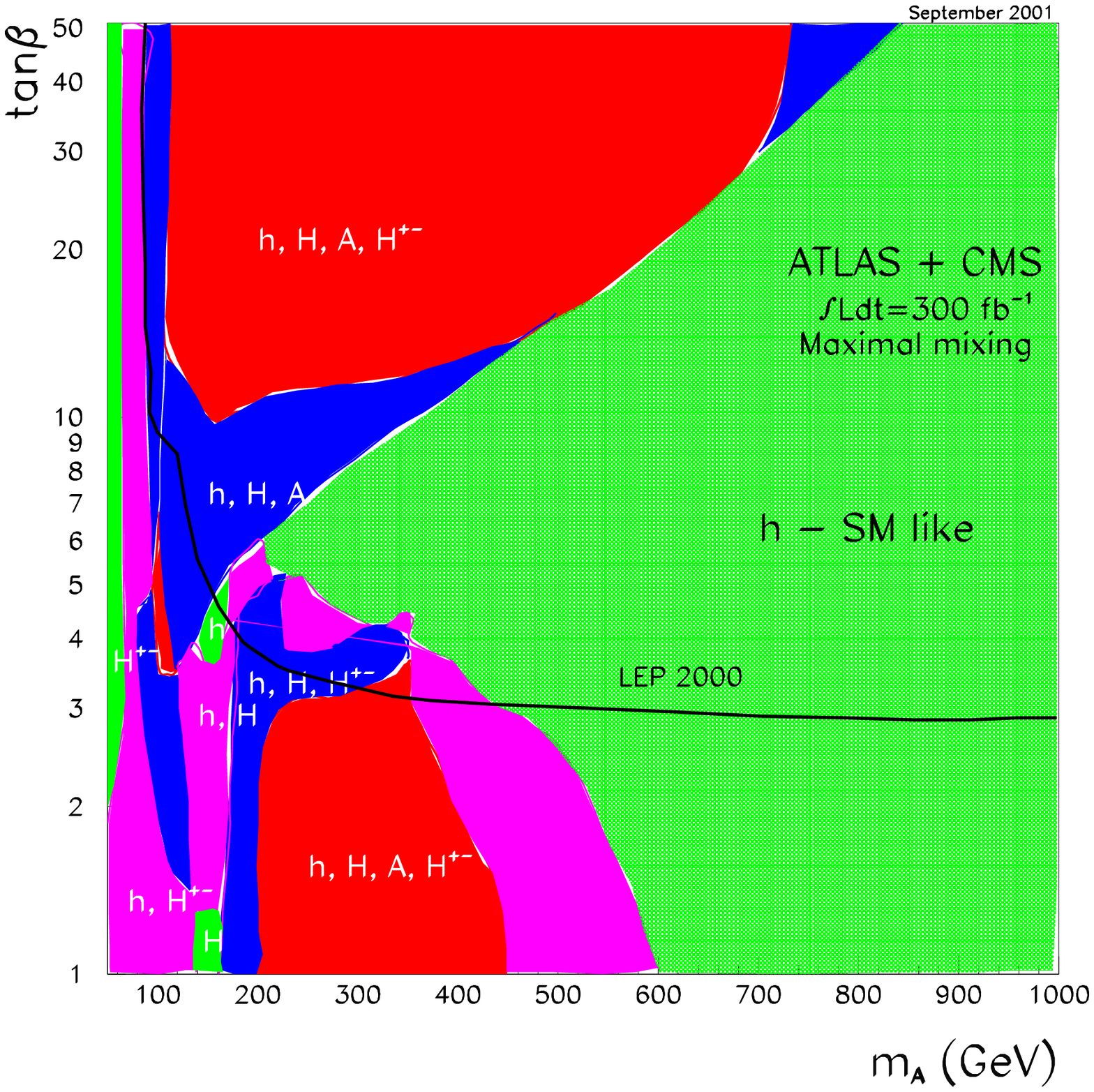}
}
\end{center}
\capt{\label{fig:susyhiggsathadron}
(a)~$5\sigma$ discovery contours for MSSM Higgs boson detection in
various channels in the $m_A$--$\tan\beta$ plane,
in the maximal mixing scenario, assuming
an integrated luminosity of $L=100~{\rm fb}^{-1}$
for the CMS detector~\protect\cite{cmsnote}. 
(b) Regions in the $\mha$--$\tan\beta$ plane in the maximal mixing scenario
in which up to four Higgs boson states
of the MSSM can be discovered at the LHC with
300~fb$^{-1}$ of data, based on a simulation that combines
data from the ATLAS and CMS detectors~\protect\cite{lhcupgrade}.}
\end{figure}

If no Higgs boson is discovered at the Tevatron, the LHC will cover the
remaining unexplored regions of the $\mha$--$\tanb$ plane, 
as shown in \fig{fig:susyhiggsathadron}~\cite{cmsnote,lhcupgrade}.
That is, in the maximal mixing scenario (and probably in most regions
of MSSM Higgs parameter space),
at least one of the Higgs bosons 
is guaranteed to be discovered at the LHC.
A large fraction of the
parameter space can be covered in the search for
a neutral CP-even Higgs boson by employing the SM Higgs search
techniques, where the SM Higgs boson is replaced by $\hl$ or
$\hh$ with the appropriate rescaling of the couplings.  Moreover,
\fig{fig:susyhiggsathadron} illustrates that
in some regions of the parameter space, both $\hl$ and $\hh$ can be
simultaneously observed, and additional Higgs search
techniques can be employed to discover $\ha$ and/or $\hpm$.

Thus, it may be possible 
at the LHC to either {\it exclude} the entire $\mha$--$\tan\beta$ plane 
(thereby eliminating the MSSM Higgs sector as a viable model), or 
achieve a $5\sigma$ discovery of at least one of the MSSM Higgs bosons,
independently of the value of $\tanb$ and $\mha$.
Note that over a significant fraction of the MSSM Higgs parameter space, 
there is still a sizable wedge-shaped
region at moderate values of $\tan\beta$,
opening up from about $\mha=200$~GeV to higher values, in which
the heavier Higgs bosons cannot be discovered at the LHC.
In this parameter regime,
only the lightest CP-even Higgs boson can be discovered, and its
properties
are nearly indistinguishable from those of the SM Higgs
boson. Precision measurements of Higgs branching ratios and other
properties will then be required in order to detect deviations from
SM Higgs predictions and
demonstrate the existence of a non-minimal Higgs sector.

For high precision Higgs measurements, we turn our attention to the
LC.  The main production mechanisms for the MSSM Higgs bosons
are~\cite{MSSMprod,Tesla-TDR,vankooten} 
\beqa &&e^+e^-\to Z\hl\,, Z\hh~~~\mbox{\rm
via Higgs-strahlung}\,,\nonumber \\ 
&& e^+e^- \to \nu
\bar{\nu}\hl\,,\nu\bar\nu\hh~~~{\rm via}~W^+W^-~{\rm
fusion}\,,\nonumber \\ 
&&e^+e^-\to\hl\ha\,, \hh\ha~~~{\rm
via}~s\mbox{\rm-channel}~Z~{\rm exchange}\,,\nonumber \\
&&e^+e^-\to H^+H^-~~~{\rm
via}~s\mbox{\rm-channel}~\gamma\,,Z~{\rm exchange}\,.  
\eeqa 
If $\sqrt{s}<2\mha$, then only $\hl$ production will be observable at the
LC.\footnote{Although $\hl\ha$ production may still be kinematically
allowed in this region, the cross-section is suppressed by a factor of
$\cos^2(\beta-\alpha)$ and is hence unobservable.}  Moreover, this
region is deep within the decoupling regime, where it will be
particularly challenging to distinguish $\hl$ from the SM Higgs boson.
%In this case, one must resort to precision measurements to distinguish
%$\hl$ from the SM Higgs boson.
As noted in Table~\ref{tab:BRmeas}, 
recent simulations of Higgs branching ratio
measurements~\cite{BattagliaDesch} suggest that the Higgs couplings to
vector bosons and the third generation fermions can be determined with
an accuracy in the range of 1--3\% at the LC.  In the approach to the 
decoupling limit, the fractional deviations of the couplings of $\hl$
relative to those of $\hsm$ scale as $m_Z^2/m_A^2$.  Thus, if
precision measurements reveal a significant deviation from SM
expectations, one could in principle derive a constraint ({\it e.g.},
upper and lower bounds) on the heavy Higgs masses.

\begin{figure}[!t]
\begin{center}
\resizebox{0.95\textwidth}{!}{
\includegraphics*[19,142][529,682]{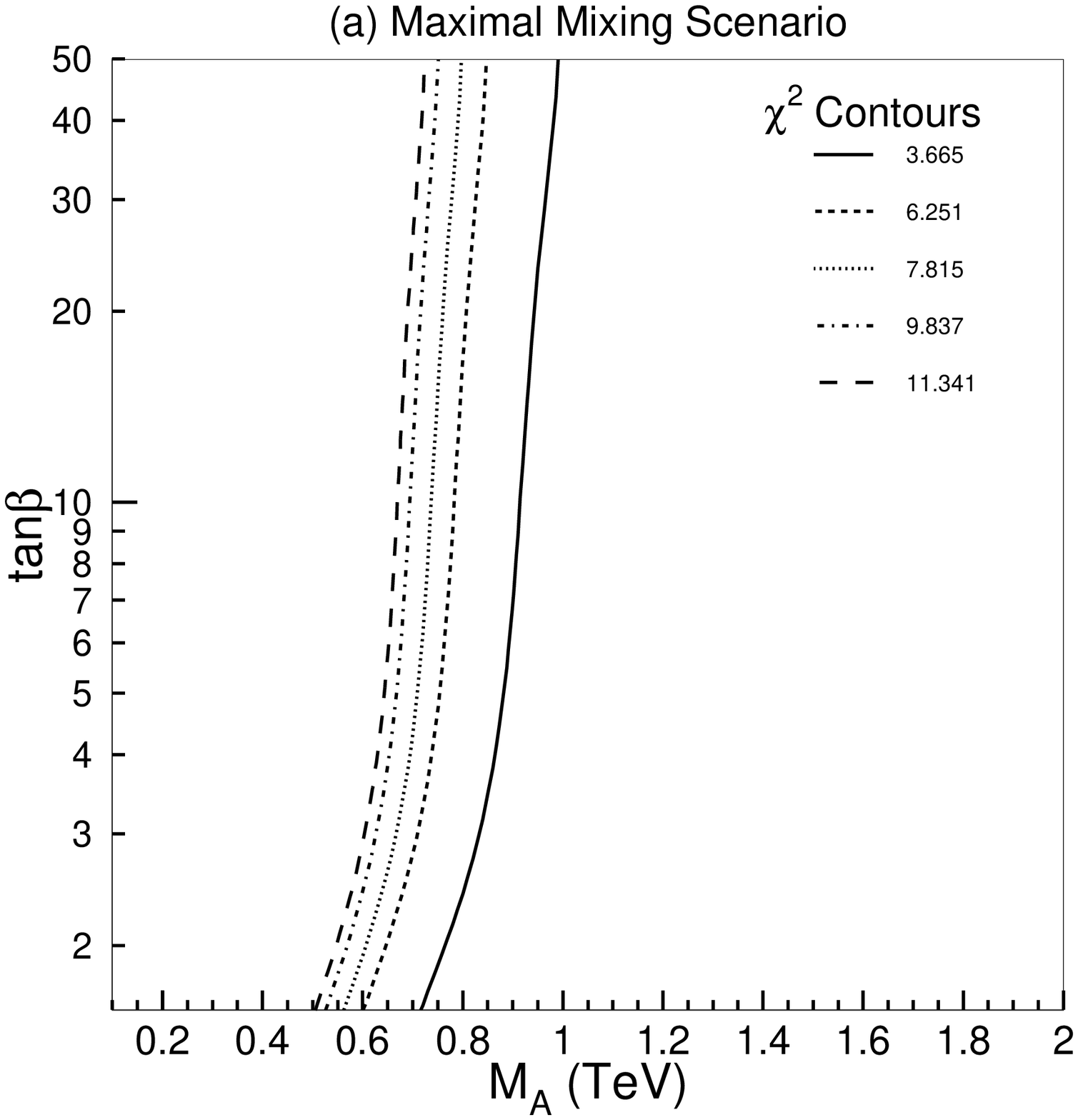}
\hspace*{5mm}
\includegraphics*[19,142][529,682]{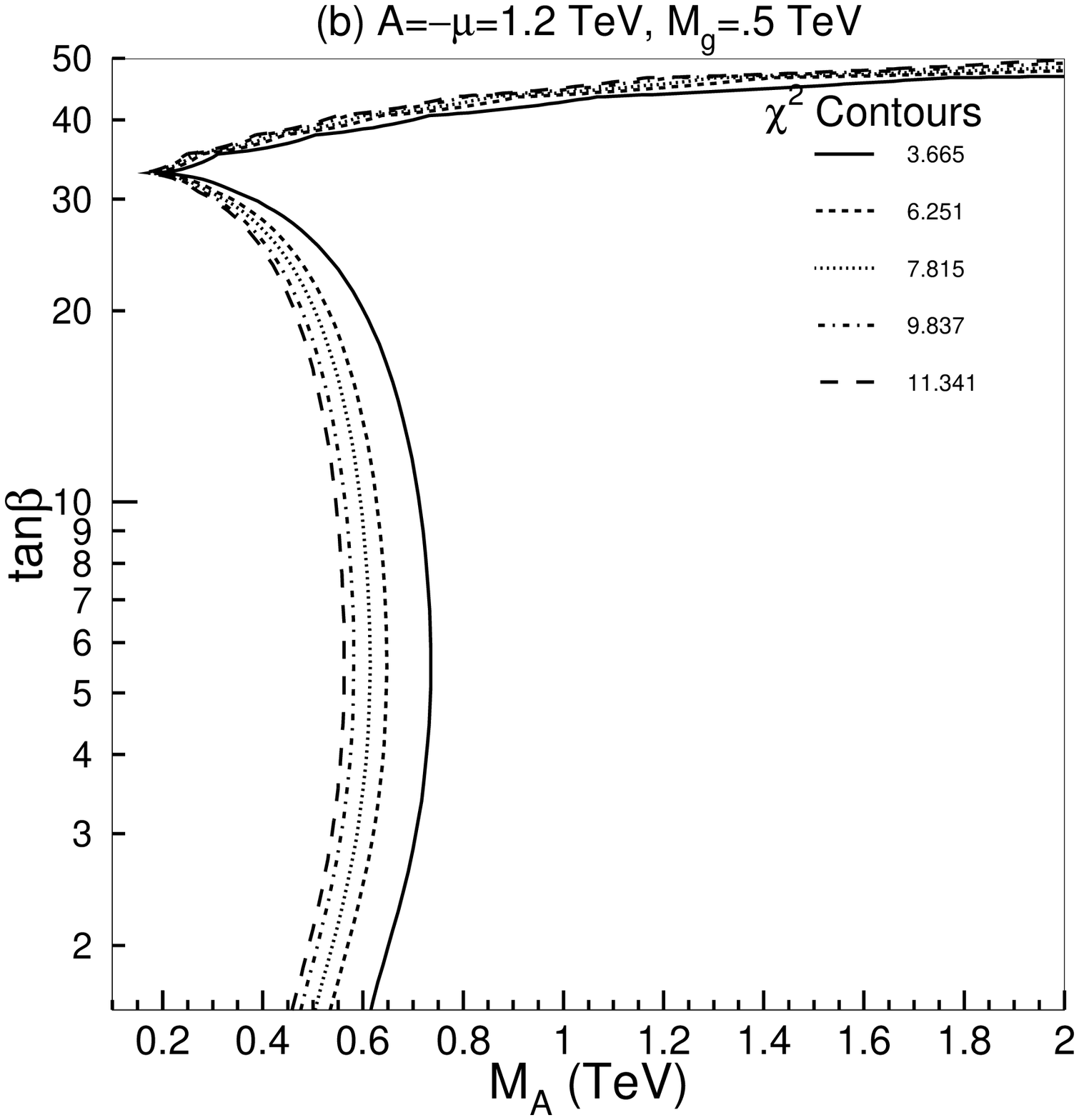}
}
\vspace{-0.15in}
\end{center}
\capt{\label{fig:chisquare}
Contours of $\chi^2$ for Higgs
boson decay observables for (a) the maximal mixing scenario; and
(b) a choice of MSSM parameters for which the loop-corrected
$\hl b\bar b$ coupling is suppressed at large $\tanb$ and low $\mha$
(relative to the corresponding tree-level coupling).
%These results are based on Higgs partial width
%measurements anticipated at the LC with
%$\sqrt{s}=500$~GeV and an integrated luminosity of 500~fb$^{-1}$.
The contours correspond to
68, 90, 95, 98 and 99\% confidence levels (right to left) for the 
observables $g^2_{hbb}$, $g^2_{h\tau\tau}$, and $g^2_{hgg}$. See
\Ref{chlm} for additional details.}
\end{figure}

In the MSSM, this constraint is
sensitive to the supersymmetric parameters that control the radiative
corrections to the Higgs couplings.  This is illustrated in
\fig{fig:chisquare}, where the constraints on $\mha$ are derived
for two different sets of MSSM parameter 
choices~\cite{chlm}.  Here, a simulation of a
global fit of measured $hbb$, $h\tau\tau$ and $hgg$ couplings is made
(based on the anticipated experimental accuracies given 
in Table~\ref{tab:BRmeas})
and $\chi^2$ contours are plotted indicating the constraints in the
$m_A$--$\tan\beta$ plane, assuming that a deviation from SM Higgs
boson couplings is seen.
In the maximal mixing scenario shown in \fig{fig:chisquare}(a),
the constraints on $\mha$ are significant and rather insensitive to the
value of $\tan\beta$. 
However in some cases, as shown in \fig{fig:chisquare}(b),
a region of $\tan\beta$ may yield almost no constraint on $\mha$.
This corresponds to the value of $\tanb$ given by
\eq{earlydecoupling}, and is a result of $\cosbma\simeq 0$ generated by
radiative corrections [$c\simeq 0$ in \eq{cbmashift}].
Thus, one cannot extract a fully model-independent upper bound on
the value of $\mha$ beyond the kinematical limit that would be obtained
if direct $\ha$ production were not observed at the LC.

\section{Conclusions}

Precision electroweak data suggest the existence of a weakly-coupled
Higgs boson.  Once the Higgs boson is discovered, one must 
determine whether it is the SM Higgs boson, or whether there are
any departures from SM Higgs predictions.  Such departures will
reveal crucial information about the nature of the 
electroweak symmetry breaking dynamics.
Precision Higgs measurements are essential for
detecting deviations from SM predictions of branching ratios, coupling
strengths, cross-sections, {\it etc.} and can provide critical tests
of the supersymmetric interpretation of new physics beyond the
Standard Model.  A program of precision
Higgs measurements will begin at the LHC, but will only truly blossom
at a future high energy $e^+e^-$ linear collider.
 
%The MSSM Higgs sector provides a potentially rich 
%phenomenology to be deciphered at future colliders.
The decoupling limit corresponds to the parameter regime in which 
the properties of the lightest CP-even Higgs boson are nearly
indistinguishable from those of the SM Higgs boson, and all other
Higgs scalars of the model are significantly heavier than the $Z$.
Deviations from the decoupling limit may provide significant information
about the non-minimal Higgs sector and can yield indirect information
about the MSSM parameters.  At large $\tanb$, there can be additional
sensitivity to MSSM parameters via enhanced radiative corrections.  It
is possible that more than one Higgs boson is accessible to future
colliders, in which case there will be many Higgs boson observables to
measure and interpret.  In contrast, the decoupling limit presents a
severe challenge for future Higgs studies and places strong
requirements on the level of precision needed to fully explore the
dynamics of electroweak symmetry breaking.

\section*{Acknowledgments}

Much of this work is based on a collaboration with Marcela Carena.
I have greatly benefited from our many animated discussions.
I am also grateful to Peter Zerwas for his kind hospitality and
his contributions to our joint efforts at the Snowmass 2001 workshop.
I am also pleased to acknowledge the collaboration with Jack
Gunion, which contributed much to my understanding of the decoupling
limit and its implications.  Finally, I would like to thank
Heather Logan and Steve Mrenna for many fruitful interactions.

This work is supported in part by the U.S. Department of Energy
under grant no.~DE-FG03-92ER40689.

%%%%%%%%%%%%%%%%%%%%%%%%%%%%%%%%%%%%%%%%%%%%%%%%%%%%%%%%%%%%%%%%%%%

\end{document}